\documentclass[showpacs,amssymb,preprint,preprintnumbers,nofootinbib,superscriptaddress]{revtex4-2}
\usepackage{amsmath}
\usepackage{graphicx}
\graphicspath{{Figures/}}
\usepackage{latexsym}
\usepackage{amsfonts}
\usepackage{url}
\usepackage{hyperref}
\hypersetup{colorlinks=true,citecolor=blue,linkcolor=blue,urlcolor=blue}

\usepackage{feynmp-auto}
\usepackage{feynmp}
 \usepackage{amssymb}
\usepackage{bm}
\usepackage{soul}
\usepackage{textcomp}
\usepackage{color}
\usepackage{bbm}
\usepackage{slashed}
\usepackage{caption}
\usepackage{epstopdf}
\usepackage{subcaption}
\usepackage{placeins}
\usepackage{color}
\usepackage{cancel} 
\usepackage[normalem]{ulem}
\usepackage{tcolorbox}
\usepackage{soul}
\usepackage{cleveref}
\usepackage{graphicx}      % For including graphics
\usepackage{subcaption}    % For subfigures
\usepackage{placeins}      % For the \FloatBarrier command
\usepackage{multirow}
\usepackage{array}

\newcommand{\ba}{\begin{eqnarray}}
\newcommand{\ea}{\end{eqnarray}}

\begin{document}

\title{Heavy-Quark Spin Symmetry Violation effects in Charmed Baryon Production}

\author{Nantana Monkata} 
\email{nantana.mon@kkumail.com}
\affiliation{Khon Kaen Particle Physics and Cosmology Theory Group (KKPaCT), Department of Physics, Faculty of Science, Khon Kaen University, 123 Mitraphap Rd., Khon Kaen, 40002, Thailand}

\author{Prin Sawasdipol}
\email{p.namwongsa@kkumail.com}
\affiliation{Khon Kaen Particle Physics and Cosmology Theory Group (KKPaCT), Department of Physics, Faculty of Science, Khon Kaen University, 123 Mitraphap Rd., Khon Kaen, 40002, Thailand}

\author{Nongnapat Ponkhuha} 
\email{nongnapat.po@kkumail.com}
\affiliation{Khon Kaen Particle Physics and Cosmology Theory Group (KKPaCT), Department of Physics, Faculty of Science, Khon Kaen University, 123 Mitraphap Rd., Khon Kaen, 40002, Thailand}

\author{Ratirat Suntharawirat} 
\email{ratirat\_su@kkumail.com}
\affiliation{Khon Kaen Particle Physics and Cosmology Theory Group (KKPaCT), Department of Physics, Faculty of Science, Khon Kaen University, 123 Mitraphap Rd., Khon Kaen, 40002, Thailand}

\author{Ahmad Jafar Arifi}
\email{ahmad.arifi@riken.jp}

\affiliation{Few-body Systems in Physics Laboratory, RIKEN Nishina Center, Wako 351-0198, Japan}
\affiliation{Research Center for Nuclear Physics (RCNP), The University of Osaka, Ibaraki, Osaka 567-0047, Japan}

\author{Daris Samart}
\email[Corresponding author: ]
{ darisa@kku.ac.th}
\affiliation{Khon Kaen Particle Physics and Cosmology Theory Group (KKPaCT), Department of Physics, Faculty of Science, Khon Kaen University, 123 Mitraphap Rd., Khon Kaen, 40002, Thailand}
%%%%%%%%%%%%%%%%%%%%%%%%%%%%%%%%%%%%%%%%%%%%%%%%%%%%%%%%%%%%%%%%%%%%%%%%%%%%%%%%%%%%%%%%%%%%%%%%%%%%%%%%%%%%%%%%%%%%%%%%%%%%%%%%%%%%%%%%%
%%%%%%%%%%%%%%%%%%%%%%%%%%%%%%%%%%%%%%%%%%%%%%%%%%%%%%%%%%%%%%%%%%%%%%%%%%%%%%%%%%%%%%%%%%%%%%%%%%%%%%%%%%%%%%%%%%%%%%%%%%%%%%%%%%%%%%%%%

\begin{abstract}
In this work, we investigate the Heavy-Quark Spin Symmetry (HQSS) exhibited in the effective Lagrangians governing the three-point interactions of $D$ mesons, charmed baryons, and nucleons. We first construct the effective Lagrangians, and there are 12 distinct terms. As a result, we observe that the invariant Lagrangian under HQSS manifests exclusively in the pseudoscalar $D$ mesons coupling to nucleons and $\Lambda_c$ baryons, whereas nucleons and $\Sigma_c$ ($\Sigma_c^*$) baryons only couple with vector $D$ mesons. By taking into account the violated heavy-quark spin transformation, one can recover all interactions from the effective Lagrangians. Furthermore, we compute the differential cross-sections of the $p\bar p \to Y_c\bar{Y}_c'$ scatterings, where $Y_c,\bar{Y_c}' = \Lambda_c,~\Sigma_c,~\Sigma_c^*$, to reveal the residue of the violating HQSS (VHQSS) on charmed baryon production. 
Ultimately, by accounting for VHQSS, we aim for precise predictions of production rates, which are essential for the High-Energy Storage Ring (HESR) experiments at the Facility for Antiproton and Ion Research (FAIR).
\end{abstract}
\maketitle{}

\newpage
\section{Introduction}
One of the challenging questions in particle physics is how strong interactions bind quarks and gluons as described by the non-abelian gauge group $SU(3)$, as is widely known quantum chromodynamics (QCD). At high energies, it is well-defined through perturbation theory due to the interaction weakening with the small couplings. However, non-perturbative QCD becomes strongly coupled and ambiguous at lower energies like a phenomenon known as color confinement. To address the questions, exploring non-perturbative phenomena, the study of heavy hadrons that contain a charm quark exhibits unique properties distinct from those of light-flavored hadrons, providing additional insights into the complexities of QCD.

The discovery of $J/\psi$ in 1974 \cite{E598:1974sol} marked the beginning of intensive research into the production of charmonium ($c\bar{c}$) states \cite{Cacciari:1995yt,Katkov:2003gj,Santoro:2013pxa,Bracko:2017fdy,Ping:2013qca}. Charmed baryon states were initially identified in 1975 during interactions with neutrinos \cite{Cazzoli:1975et}. Since then, various facilities such as CLEO \cite{Chen:2014lga}, BABAR \cite{BaBar:2004oro,BaBar:2005hhc}, Belle \cite{Belle:2005lik,Belle:2003nnu,Belle:2007hrb}, BESIII \cite{Balossino:2019nha,Liu:2023hhl}, LHCb collaborations \cite{LHCb:2013kgk,LHCb:2014zfx} have observed various hadrons that are commonly known as exotic hadrons.  Nevertheless, there has been less comprehensive research into the production and spectroscopy of charmed baryons compared to that of charmonium states despite their ability to offer similar information about the quark confinement mechanism. In the near future, FAIR at GSI, specifically the $\bar{\text{P}}$ANDA (Antiproton Annihilation at Darmstadt), will aim for highly accurate spectroscopy of charmed hadrons and exploration of their interactions with ordinary matter \cite{PANDA:2009yku}.

Several studies have calculated cross-sections for the production of $p\bar{p} \rightarrow \Lambda_c\bar{\Lambda}_c $. The Quark-Gluon String Model (QGSM) and Regge approach were applied in Refs.~\cite{Kaidalov:1994mda,Titov:2008yf,Khodjamirian:2011sp}. In these studies, the annihilation of an initial pair $q\bar{q}$ produces charmed hadrons through string fission. In Refs.~\cite{Haidenbauer:2010nx,Haidenbauer:2009ad,Haidenbauer:2016pva}, a meson exchange framework was applied and also the production rate of the charmed baryon $\Lambda_{c}^{}$(2940) was estimated in the $p\bar{p}$  annihilation at $\bar{\text{P}}$ANDA energies \cite{He:2011jp}.  The effective Lagrangian model was calculated in Refs.~\cite{Shyam:2014dia,Sangkhakrit:2020wyi} within a single channel reaction that described the reaction as the sum of the $t$-channel $D^{0}$ and $D^{0*}$ meson-exchange processes. In previous studies, the predicted production vary depending on the model used, it revealed a strongly model dependent as their charm production differ from each other by several orders, and there is no agreement on the best way to describe this reaction.

Heavy-quark spin symmetry (HQSS) plays a significant role in understanding of low-energy strong interactions and the classification of the heavy-light hadronic spectrum including the dynamics charm baryons as studied in Refs.~\cite{Du:2022rbf,Baru:2022xne,Samart:2016doe,Baru:2017qwx,Romanets:2012ce}. In the infinite heavy-quark mass limit ($m\rightarrow\infty$), the degrees of freedom (DOF) associated with the heavy quark decouple from those of the light quark. The HQSS implies that the pseudoscalar $D$ meson ($J^{P}=0^{-}$) and the vector $D$ meson ($J^{P}=1^{-}$), as well as charm baryon sextets such as  $\Sigma_{c},$ $\Xi_{c}$ and $\Omega_{c}$ with spin $1/2$ ($J^{P}=1/2^{+}$)  and $\Sigma^{*}_{c},$ $\Xi_{c}^{*}$ and $\Omega_{c}^{*}$ with spin $3/2$ ($J^{P}=3/2^{+}$), form degenerate states \cite{Lutz:2011fe,Suh:2022ean,Kim:2018nqf}, which depend on the spin flip of the charm quark. Although HQSS is primarily effective in describing on-shell quark behaviors, in our process, the charm quark is significantly off-shell due to interactions mediated by virtual $D$ mesons that do not obey the on-shell mass-energy relation \cite{Hussain:1994zr,alexander2014}.

This study focuses on investigating charm production processes such as \( p\bar{p} \rightarrow \Lambda_{c}\bar{\Lambda}_{c}, \Sigma_{c}\bar{\Sigma}_{c},\) \( \Sigma_{c}^{*}\bar{\Sigma}_{c}^{*}, 
\Sigma_{c}^{*}\bar{\Sigma}_{c},\Sigma_{c}\bar{\Sigma}_{c}^{*}, \Sigma_{c}\bar{\Lambda}_{c}, \Lambda_{c}\bar{\Sigma}_{c}, \Sigma_{c}^{*}\bar{\Lambda}_{c}, \Lambda_{c}\bar{\Sigma}_{c}^{*} \) using the effective Lagrangians approach with $SU(2)_f$ symmetry. This investigation aims to examine the consequences of HQSS and its violations by combining these constraints to evaluate their effects on scattering processes. We acknowledge the significant role of HQSS in improving the accuracy of our theoretical framework. Additionally, we estimate various coupling constants under $SU(4)_f$ symmetry breaking, assuming a deviation 20\% relative to $SU(3)_f$ \cite{Skoupil:2020tge}. Ultimately, we aim for precise predictions of the cross sections with beam momenta ($p_{\text{Lab}}$) ranging from the threshold to 15 GeV/c. This range covers the beam momenta of interest for the $\bar{\text{P}}$ANDA experiment at the Facility for Antiproton and Ion Research (FAIR) \cite{PANDA:2009yku}, and we also seek to refine the effective field theory (EFT) of heavy hadrons.

The present work is organized as follows; in the section \ref{sec-2}, we set up the effective Lagrangians in terms of $SU(2)_f$ symmetry and construct the conserving and violating HQSS Lagrangians with their implications in the $SU(2)_f$ effective Lagrangians.  In the next section, we will compute the scattering amplitudes and the differential cross-sections in our model. In the section \ref{sec-4}, the numerical results of all relevant observables for the charmed baryon productions are presented. Finally, we close this work with discussions and conclusions in the section \ref{sec-5}.  

\section{Formalism}\label{sec-2}
%{The $SU(2)_f^{}$ Effective Lagrangians}
Firstly, to construct the $SU(2)_f$ effective Lagrangians of $D$ mesons, nucleon, and charmed baryons. %at chiral power counting $Q^{0}$ and $Q^{1}$. 
In addition, all relevant symmetries of the effective Lagrangians in the system are considered. The basic building blocks of the charmed baryon productions from $p\bar p$ scattering are introduced as 
\begin{align}
N_{a}, \quad \Lambda_{(c)}, \quad\Sigma_{(c)ab}, \quad\Sigma_{(c)ab}^{\mu},\quad D_{a}, \quad D_{\mu a},
\end{align}
where the light baryon singlet fields 
$N_{a}(J^P=\frac{1}{2}^{+})$, charmed baryon singlet fields $\Lambda_{(c)}(J^P=\frac{1}{2}^{+})$ and charmed baryon triplet fields $\Sigma_{(c)ab}(J^P=\frac{1}{2}^{+})$ and $\Sigma_{(c)ab}^{\mu}(J^P=\frac{3}{2}^{+})$. In addition, the Latin indices, $a,b,\cdots=1,2$ are the fundamental indices of the $SU(2)_f$ symmetry for the hadronic fields in this work.  The $D$ mesons fields $D(J^P=0^{+})$ and $D_\mu(J^P=1^{+})$ are pseudoscalar and vector $D$ mesons respectively, they represent forming a doublet. Furthermore, the nucleon, charmed baryons and, $D$ mesons can be explicitly represented in the $SU(2)_f$ space by
\begin{align}
&N_{a}=\binom{p}{n}_{a}, \quad
\Lambda_{(c)}=\Lambda_{(c)}^{+},\quad D_{a}=\begin{pmatrix}D^0 & D^{+}\end{pmatrix}_{a},  \quad D_{\mu}=\begin{pmatrix}D^{0}_{\mu} & D^{+}_{\mu}\end{pmatrix}_{a},\\
&\Sigma_{(c)ab}  
=\left(\begin{array}{cc}
\frac{1}{\sqrt{2}} \Sigma_c^{+} & \Sigma_c^{++} \\
\Sigma_c^0 & -\frac{1}{\sqrt{2}} \Sigma_c^{+}
\end{array}\right)_{ab},~\Sigma_{(c)ab}^{\mu}  
=\left(\begin{array}{cc}
\frac{1}{\sqrt{2}} \Sigma_c^{\mu,+} & \Sigma_c^{\mu,++} \\
\Sigma_c^{\mu,0} & -\frac{1}{\sqrt{2}} \Sigma_c^{\mu,+}
\end{array}\right)_{ab},
\label{eq:sigmaep}
\end{align}
 Using these building blocks, the $SU(2)_f$ effective Lagrangians of $D$ mesons, light, and charmed baryons 
%at chiral power counting $Q^0$ and $Q^{1}$ 
are given by
\begin{align}
\mathcal{L} &= \mathcal{L}^{P} + \mathcal{L}^{A} + \mathcal{L}^{V} + \mathcal{L}^{T}, 
\\
\mathcal{L}^P &= g_1^{} \bar{\Sigma}_{(c)ab}^{} i \gamma_5^{} N_b^{} D_a^{} + g_2^{} \bar{\Lambda}_{(c)} i \gamma_5^{} N_a^{} D_a^{} + \mathrm{h.c.}, 
\label{mathcal{L}^P}
\\
\mathcal{L}^A 
&= -\frac{g_3^{}}{m_D^{}} \bar{\Sigma}_{(c)ab}^{} \gamma^{\mu} \gamma_5^{} N_b^{} \partial_{\mu}^{} D_a^{} -\frac{g_4^{}}{m_D^{}} \bar{\Lambda}_{(c)} \gamma^\mu \gamma_5^{} N_a^{} \partial_\mu^{} D_a^{} - \frac{g_5^{}}{m_D^{}} \bar{\Sigma}^\mu_{ab} N_b^{} \partial_\mu^{} D_a^{} + \mathrm{h.c.},
\label{mathcal{L}^A}
\\
\mathcal{L}^V 
&= f_1^{} \bar{\Sigma}_{(c)ab}^{} \gamma^\mu N_b^{} D_{\mu a} + f_2^{} \bar{\Lambda}_{(c)} \gamma^\mu N_a^{} D_{\mu a}^{} + f_3^{} \bar{\Sigma}^\mu_{ab} \gamma_5^{} N_b^{} D_{\mu a} + \mathrm{h.c.},
\label{mathcal{L}^V}\\
\mathcal{L}^T &= \frac{h_1^{}}{2m_{D^{*}}} \bar{\Sigma}_{ab} \sigma^{\mu\nu} N_b^{} D_{\mu \nu}^a + \frac{h_2^{}}{2m_{D^{*}}} \bar{\Lambda}_{(c)} \sigma^{\mu \nu} N_a^{} D_{\mu \nu}^a -i \frac{h_3^{}}{2m_{D^{*}}} \bar{\Sigma}^{\mu}_{(c)ab} \gamma^{\nu} \gamma_5^{} N_b^{} D_{\mu \nu}^a \nonumber\\
&\quad + \frac{h_4^{}}{4m_{D^{*}}} \epsilon_{\mu \nu \alpha \beta}^{} \bar{\Sigma}^{\mu}_{(c)ab} \gamma^{\nu} N_{b}^{} D^{\alpha \beta}_{a} + \mathrm{h.c.}, \label{mathcal{L}^T}
\end{align}
where $D_{\mu\nu} \equiv \partial_\mu D_\nu - \partial_\nu D_\mu$ in $\mathcal{L}^{T}$ is the second rank tensor $D_{\mu\nu}$ corresponds to the partial derivative of vector $D$ meson fields. %The Latin alphabets are the fundamental indices of the $SU(2)_f$ flavor symmetry i.e., $a,~b,\cdots = 1,~2$\,. 

Notice that there are 12 low-energy constants (LECs) in the $SU(2)_f$ effective Lagrangians above where $g$, $f$, and $h$ represent the pseudoscalar, axial-vector, vector, and tensor couplings.

%%%%%%%%%%%%%%%%%%%%%%%%%%%%%%%%%%%%%%%%%%%%%%%%%%%%%%%%%%%%%%%%%%%%%%%%%%%%%%%%%%%%%%%%%%%%%%%%%%%%%%%%%%%%%%%%%%%%%%%%%%%%%%%%%%%%%%%%%%%%%%%%%%%%%%%%%%%%
\subsection{Super-multiplet fields in Heavy-Quark Spin Symmetry}
In heavy-quark spin symmetry (HQSS) in QCD, we introduce as slowly varying fields, $D_\pm(x)$\,, $D^\mu_\pm(x)$\,, $\Sigma_{(c,\pm)}(x)$\,, $\Sigma_{(c,\pm)}^\mu(x)$ and $\Lambda_{(c,\pm)}$ (we drop the flavor indices here for simplicity). $D$ mesons and charmed baryons fields are decomposed as the following expressions,
\begin{align}
D(x)\:\, &= e^{-i\,(v\cdot x) \,M_c}\,D_{+}(x) + e^{+i\,(v\cdot x) \,M_c}\,D_{-}(x)\,,\nonumber\\
D^{\mu}(x) &= e^{-i\,(v\cdot x) \,M_c}\,D^\mu_{+}(x) + e^{+i\,(v\cdot x) \,M_c}\,D^\mu_{-}(x)\,,\nonumber\\
\Sigma_{c}(x)\: &= e^{-i\,(v\cdot x) \,M_{\Sigma_c}}\,\Sigma_{c,+}(x) +e^{+i\,(v\cdot x) \,M_{\Sigma_c}}\,\Sigma_{c,-}(x)\,,\nonumber\\
\Sigma_{c}^\mu(x)\;\! &= e^{-i\,(v\cdot x) \,M_{\Sigma_c^*}}\,\Sigma_{c,+}^\mu(x) +e^{+i\,(v\cdot x) \,M_{\Sigma_c^*}}\,\Sigma_{c,-}^\mu(x)\,,\nonumber\\
\Lambda_{c}(x)\: &= e^{-i\,(v\cdot x) \,M_{\Lambda_c}}\,\Lambda_{c,+}(x) +e^{+i\,(v\cdot x) \,M_{\Lambda_c}}\,\Lambda_{c,-}(x)\,,
\label{non-relativistic-expansion}
\end{align}
with a 4-velocity $v$ normalized by $v^2=1$. The mass parameters $M_c$, $M_{\Sigma_c}$, $M_{\Sigma_c^*}$, and $M_{\Lambda_c}$ are the charm quark, $\Sigma_c$ spin-$1/2$, $\Sigma_c^*$ spin-$3/2$, and $\Lambda_c$ masses respectively.
In the heavy-quark limit $m_Q \rightarrow \infty$, the spin interaction between light and heavy quarks has disappeared. 
As a consequence, the pseudoscalar and vector $D$ mesons as well as spin-$\frac{1}{2}$ and spin-$\frac{3}{2}$ baryons form degenerate states, which can be defined by
\begin{align}
H_{a} & =\left(\frac{1+\slashed{v}}{2}\right)\left( D_{\mu a,+}\gamma^{\mu} + i \gamma_5 D_{a,+} \right) ,
\label{H-D-meson}\\
T^{\mu}_{ab} & =\frac{1}{\sqrt{3}}\left(\gamma^{\mu}+v^{\mu}\right)i \gamma_{5}\left(\frac{1+\slashed{v}}{2}\right) \Sigma_{(c,+)ab}+\left(\frac{1+\slashed{v}}{2}\right) \Sigma_{(c,+)ab}^\mu \,,
\label{H-S-baryon}\\
T & =\frac{1+\slashed{v}}{2} \Lambda_{(c,+)}\,,
\label{H-T-baryon}
\end{align}
and their conjugate fields are
\begin{align}
\bar{H}_{a} =\gamma_0 H^{\dagger}_{a} \gamma_0\,,
\quad
\bar{T}^{\mu}_{ab} =\left(T^{\mu}_{ab}\right)^{\dagger} \gamma_0 \,,\quad 
\bar{T}  =\left(T\right)^{\dagger} \gamma_0\,.
\end{align}
These super-multiplet heavy-quark hadronic fields are building blocks for the HQSS Lagrangian. These building blocks obey the following $S U(2)_v$ transformations,
\begin{align}
H_{a} & \rightarrow e^{-i\theta_{\alpha}S^{\alpha}} H_{a}\,,
\qquad
\bar{H}_{a}  \rightarrow \bar{H}_{a} e^{i\theta_{\alpha}S^{\alpha}}\,,
\\
T^{\mu}_{ab} & \rightarrow e^{-i S_\alpha \theta^\alpha} T^{\mu}_{ab}\,,\qquad
\bar{T}^{\mu}_{ab}  \rightarrow \bar{T}^{\mu}_{ab} e^{i\theta_{\alpha}S^{\alpha}}\,,
\\
T & \rightarrow e^{-i S_\alpha \theta^\alpha}T\,,\qquad\quad
\bar{T} \rightarrow \bar{T} e^{i S_\alpha \theta^\alpha}\,,
\end{align}
where $S^\alpha$ is the heavy quark spin operator. We note that all super-multiplet heavy-quark hadronic fields transform as doublet under $S U(2)_v$ HQSS where as the nucleon field is transformed as a singlet under the $SU(2)_v$ symmetry. The definition of the heavy-quark spin operator and its properties are read,
\begin{align}
S^{\alpha} = \frac{1}{2} \gamma_5\left[\slashed{v}, \gamma^{\alpha}\right]\,, \quad
S_\alpha^{\dagger} \gamma_0  =\gamma_0 S_\alpha\,, \quad
\left[\slashed{v}, S_\alpha\right]  =0\,, \quad \left[S_\alpha, \gamma_{5} \right] = 0\,.
\end{align}
Taking into account for all of the super-multiplet fields of the HQSS, the conserving HQSS (CHQSS) Lagrangian is given by
\begin{align}
\mathcal{L}_{\rm CHQSS} &=  c_1^{} \left\langle \bar{T}^\mu_{ab} \gamma_5^{} N_b^{} H_a^{} \gamma_\mu^{} + \rm{h.c.} \right\rangle +  c_2 \left\langle \bar{T} \gamma_5^{} N_a^{} H_a^{} \gamma_5^{} + \rm{h.c.} \right\rangle, 
\label{HQSSC}
\end{align}
where $\langle \quad \rangle$ stands for trace over the $SU(2)_v$ space. On the other hand, the violating HQSS (VHQSS) Lagrangian is read,
\begin{align}
\mathcal{L}_{\mathrm{VHQSS}} =\,&  b_1^{} \left\langle \bar{T}^\mu_{ab} \gamma^\nu N_b^{} H_a^{}  \gamma_\mu^{} \gamma_\nu^{} \gamma_5^{} + \rm{h.c.} \right\rangle +  b_2^{} \left\langle \bar{T} \gamma^{\mu} N_a^{} H_a^{} \gamma_\mu + \rm {h.c.} \right\rangle + b_3^{} \left\langle \bar{T} \sigma^{\mu\nu}  N_a^{} H_a^{} \sigma_{\mu\nu}^{} + \rm{h.c.} \right\rangle.
\label{HQSSV}
\end{align}
Having used the definitions of the super-multiplet heavy-quark fields in Eqs.~(\ref{H-D-meson}-\ref{H-T-baryon}), a trace of Eqs.~(\ref{HQSSC}, \ref{HQSSV}) are given by
\begin{align}
\mathcal{L}_{\mathrm{CHQSS}} =&\;   \frac{1}{\sqrt{3}}c_1 \bar{\Sigma}_{(c)ab}^{} \gamma^\mu N_b^{} D_{\mu a} + c_1^{} \bar{\Sigma}_{(c)ab}^{\mu} \gamma_5^{} N_b^{} D_{\mu a} + i c_2^{} \bar{\Lambda}^{+}_{(c)} \gamma_5^{} N_a^{} D_a + \rm{h.c.}, 
\label{tr-LCHQSS}\\
\mathcal{L}_{\mathrm{VHQSS}} =&\; \frac{i\sqrt{3}}{2}b_1^{} \bar{\Sigma}_{(c)ab}^{} \gamma_5^{} N_b^{} D_a^{} + \frac{i}{\sqrt{3}}b_1^{}  \bar{\Sigma}_{(c)ab}^{} \sigma_{\mu \nu} N_b^{} D^{\mu}_a v^{\nu} - \frac{i }{2}b_1^{} \epsilon_{\mu \nu \alpha \beta}^{} \bar{\Sigma}_{(c)ab}^\mu \gamma^\nu N_b^{} D^\alpha_a v^\beta 
\nonumber\\
+&\;b_{2}\Bar{\Lambda}_{(c)}^{+}\gamma^{\mu}N_{a}D_{\mu a} + ib_{3}\Bar{\Lambda}_{(c)}\sigma^{\mu \nu} N_{a}D_{\mu}v_{\nu}+\rm{h.c.}\,.
\label{tr-LVHQSS}
\end{align}
It is worth discussing the implications of the \emph{conserving} and \emph{violating} Lagragians for $D$ mesons, charmed baryons, and nucleon couplings. In the conserving HQSS limit, the $\Lambda_c$ baryon exclusively couples to the pseudoscalar $D$ meson whereas the $\Sigma_c$ and $\Sigma_c^*$ couple with the vector $D$ meson only. Taking into account of the non-invariant HQSS terms, charmed baryons can couple with the pseudoscalar and vector $D$ mesons.

Then, we substitute the non-relativistic expansion for slowly varying fields from Eq.~(\ref{non-relativistic-expansion}) into the $SU(2)_f$ effective Lagrangians presented in Eqs.~(\ref{mathcal{L}^P}-\ref{mathcal{L}^T}). Next, we use Eqs.~(\ref{tr-LCHQSS}) and (\ref{tr-LVHQSS}) to identify which terms in the Lagrangian expansion are \emph{conserving} or \emph{violating} terms. The conserving terms are given by
\begin{align}
\begin{split}
\mathcal{L} =&\;  g_2^{} \bar{\Lambda}_{(c)} i \gamma_5^{} N_a^{} D_a^{} - \frac{g_4^{}}{m_D^{}} \bar{\Lambda}_{(c)} \gamma^\mu \gamma_5^{} N_a^{} \partial_\mu^{} D_a^{}\\+&\;f_1^{} \bar{\Sigma}_{(c)ab}^{} \gamma^\mu N_b^{} D_{\mu a} + f_3^{} \bar{\Sigma}^\mu_{ab} \gamma_5^{} N_b^{} D_{\mu a}+ \frac{ih_{3}}{2m_{D^{*}}}\Bar{\Sigma}^{\mu}_{(c)ab}\gamma^{\nu}\gamma_{5}N_{b}\partial_{\nu}D_{\mu a}+\rm{h.c.}\, ,
\label{SU2-CHQSS}
\end{split}
\end{align}
and the violating terms are read
\begin{align}
\begin{split}
\mathcal{L}=&\;  g_1^{} \bar{\Sigma}_{(c)ab}^{} i \gamma_5^{} N_b^{} D_a^{}
- \frac{g_3^{}}{m_D^{}} \bar{\Sigma}_{(c)ab}^{} \gamma^{\mu} \gamma_5^{} N_b^{} \partial_{\mu}^{} D_a^{}
\\+&\; f_2^{} \bar{\Lambda}_{(c)} \gamma^\mu N_a^{} D_{\mu a}^{}
+ \frac{h_{1}}{m_{D^{*}}}\Bar{\Sigma}_{(c)ab}\sigma^{\mu \nu}\partial_{\mu}N_{b}D_{\nu a}
+ \frac{h_{2}}{m_{D^{*}}}\Bar{\Lambda}_{(c)ab}\sigma^{\mu \nu}\partial_{\mu}N_{a}D_{\nu a}
\\-&\;\frac{h_{4}}{m_{D^{*}}}\epsilon_{\mu\nu\alpha\beta}\Bar{\Sigma}_{(c)ab}^{\mu}\gamma^{\nu}N_{b}\partial^{\alpha}D^{\beta}+\rm{h.c.}\,.
\label{SU2-VHQSS}
\end{split}
\end{align}
 We match the relations between effective Lagrangians of the super-multiplet heavy-quark sum rules in Eqs.~(\ref{tr-LCHQSS}, \ref{tr-LVHQSS}) and the heavy-quark mass expansion of the effective Lagrangians in Eqs.~(\ref{SU2-CHQSS}, \ref{SU2-VHQSS}). For the perfect HQSS limit, we find,
\begin{align}
\begin{split}
c_{1} = \sqrt{3}f_{1}, \quad
2c_{1} = f_{3}+\frac{1}{2}h_{3}, \quad c_{2} = g_{2} + g_{4}\,.
\label{c-sumrule}
\end{split}
\end{align}
On the other hand, the HQSS violation case gives
\begin{align}
\begin{split}
b_{1} = \frac{2}{\sqrt{3}}(g_{1} + g_{3}),\quad
b_{1} = \sqrt{3}h_{1}, \quad
b_{1} = 2h_{4}, \quad 
b_{2} = f_{2}, \quad 
b_{3} = h_{2}\,.
\label{b-sumrule}
\end{split}
\end{align}
These lead to the heavy-quark sum rules, and we find 3 sum rules as 
\begin{align}
f_{1}= \, \frac{1}{2\sqrt{3}}\Big(f_{3}+\frac{1}{2}h_{3}\Big), \quad h_{1} = \frac{2}{3}(g_{1} + g_{3}),\quad h_{4} = \frac{\sqrt{3}}{2}h_{1}\,.
\end{align}
\label{Section A}
The 3 sum rules above reduce the number of the LECs to 9 parameters.  
%%%%%%%%%%%%%%%%%%%%%%%%%%%%%%%%%%%%%%%%%%%%%%%%%%%%%%%%%%%%%%%%%%%%%%%%%%%%%%%%%%%%%%%%%%%%%%%%%%%%%%%%%%%%%%%%%%%%%%%%%%%%%%%%%%%%%%%%%%%%%%%%%%%%%%%%%%%%
\subsection{Determination of coupling constants (LECs) of the Effective Lagrangians}
Due to the limited experimental data relevant to our study, we can not fix the LECs using the data. Theoretical estimations are required to determine these LECs. In this work, we will use the $SU(4)_f$ symmetry breaking of the baryon-baryon-meson interactions. For the pseudoscalar $D$ meson coupling under $SU(4)_f$ symmetry,  one finds \cite{Haidenbauer:2016pva}
\begin{align}
&g_1\equiv g^{(P)}_{DN\Sigma_{c}}=\left(1-2 \alpha_{p s}\right) g_{N N \pi}\,,
\\
&g_2 \equiv g^{(P)}_{DN\Lambda_{c}} =-\frac{1}{\sqrt{3}}\left(1+2 \alpha_{p s}\right) g_{N N \pi} \,.
\end{align}
Given the experimental measurement of $g^{2}_{NN\pi}/4\pi=14.4$ \cite{Janssen:2003qc,Skoupil:2020tge} and $\alpha_{ps} = 2/5$, which is the ratio determined by the non-relativistic quark model with the $SU(6)$ symmetry. \\
Then, we obtain 
\begin{align}
\begin{split}
g_1=  \pm 2.69\,,\quad g_2= \pm 13.98\,.
\end{split}
\end{align}
%\clearpage
For the axial-vector ($g_{3,4}$), vector ($f_{1,2}$), and tensor couplings ($h_2$), we found that the $SU(4)_f$ symmetry of the baryon-baryon-meson couplings implies 
\begin{eqnarray}
g_3 &\equiv& g^{(A)}_{DN\Sigma_{c}} = g^{(A)}_{KN\Sigma}\,,
\nonumber\\
g_4 &\equiv& g^{(A)}_{DN\Lambda_{c}} = g^{(A)}_{KN\Lambda}\,,
\nonumber\\
f_1 &\equiv& f_{DN\Sigma_{c}} = f_{KN\Sigma}\,,
\nonumber\\
f_2 &\equiv& f_{DN\Lambda_{c}} = f_{KN\Lambda}\,,
\nonumber\\
h_2 &\equiv& h_{DN\Lambda_{c}} = h_{KN\Lambda}\,.
\end{eqnarray}
Noting that these results of the coupling constants (LECs) have been used to study the charmed baryon productions in Ref. \cite{Sangkhakrit:2020wyi} where the values of these couplings were estimated in $SU(3)_f$ symmetry. However, $SU(4)_f$ symmetry is not a good approximation to include the light and heavy quarks in the same group. Using the $^3P_0$ quark model, it has been shown that there are effects of the $SU(4)_f$ symmetry breaking in the heavy baryon, light baryon, and heavy meson couplings ~\cite{Krein:2012lra,Fontoura:2017ujf}. As a result, the flavor symmetry breaking effect provides a reasonable first approximation for this correspondence, with a 20\% symmetry breaking relative to $SU(4)_f$~\cite{Krein:2012lra,Fontoura:2017ujf,Skoupil:2020tge} values, as shown in Table \ref{Table coupling}.

\begin{table}[h!]
\centering
\renewcommand{\arraystretch}{1.4}
\setlength{\tabcolsep}{7pt} 
\begin{tabular}{| c | c |c |c| c |c| c  |} 
 \hline
  ${g_{1}}$ &${g_{2}}$ & ${g_{3}}$ & ${g_{4}}$ & ${f_{1}}$ & ${f_{2}}$ & ${h_{2}}$  \\ 
\hline \hline
 $\pm 2.69$ & $\pm 13.98$ & $-2.50\,\pm\,0.50$ & $-13.50\,\pm\,2.70$ & $-4.182\,\pm\,0.836$ & $5.11\,\pm\,1.02$ & $10.40\,\pm\,2.08$ \\
 \hline
 \end{tabular}
\caption{The coupling constants (LECs) for $D$ mesons, charmed baryons, and light baryons derived from $SU(4)_f$ symmetry breaking.}
\label{Table coupling}
\end{table}
\label{Derive coupling constants}
Next, we can calculate the scattering amplitudes by applying these coupling values. However, the other unknown parameters $f_3$,\,$h_1$,\,$h_3$ and $h_4$\, will be estimated by matching the relations described in Eqs.~(\ref{c-sumrule}, \ref{b-sumrule}). In addition, it is worth noting that we have dropped $g_5$ in the effective Lagrangian in Eq.~(\ref{mathcal{L}^A}) since this term vanishes in the heavy-quark limit. 

%%%%%%%%%%%%%%%%%%%%%%%%%%%%%%%%%%%%%%%%%%%%%%%%%%%%%%%%%%%%%%%%%%%%%%%%%%%%%%%%%%%%%%%%%%%%%%%%%%
%%%%%%%%%%%%%%%%%%%%%%%%%%%%%%%%%%%%%%%%%%%%%%%%%%%%%%%%%%%%%%%%%%%%%%%%%%%%%%%%%%%%%%%%%%%%%%%%%%
\section{Scattering Amplitudes}\label{sec-3}
%\unitlength = 1mm
%\begin{figure}[h]
 %   \centering
  %  \begin{fmffile}{simple1}
        %\begin{fmfgraph*}(40,25)
        %\fmfleft{i1,i2}
        %\fmfright{o1,o2}
        %\fmflabel{$r,p$}{i1}
        %\fmflabel{$r_{1},\bar{p}$}{i2}
        %\fmflabel{$r'_{1},Y_{(c)}$}{o1}
        %\fmflabel{$r'_{2},\Bar{Y'}_{(c)}$}{o2}
        %\fmf{fermion}{i1,v1,o1}
        %\fmf{fermion}{o2,v2,i2}
        %\fmf{dashes,label=$\phi_{c}(q)$}{v1,v2}
        %\fmfdot{v1,v2}
        %\fmflabel{$x_1$}{v1}
        %\fmflabel{$x_2$}{v2}
        %\fmf{plain,label=$p_1$}{i1,v1}
        %\fmf{plain,label=$p'_1$}{v1,o1}
        %\fmf{plain,label=$p_2$}{i2,v2}
        %\fmf{plain,label=$p'_2$}{v2,o2}
 % \end{fmfgraph*}
   % \end{fmffile}
    %\vspace{5mm}
\begin{figure}[h]
  \centering
\includegraphics[width=0.325\textwidth]{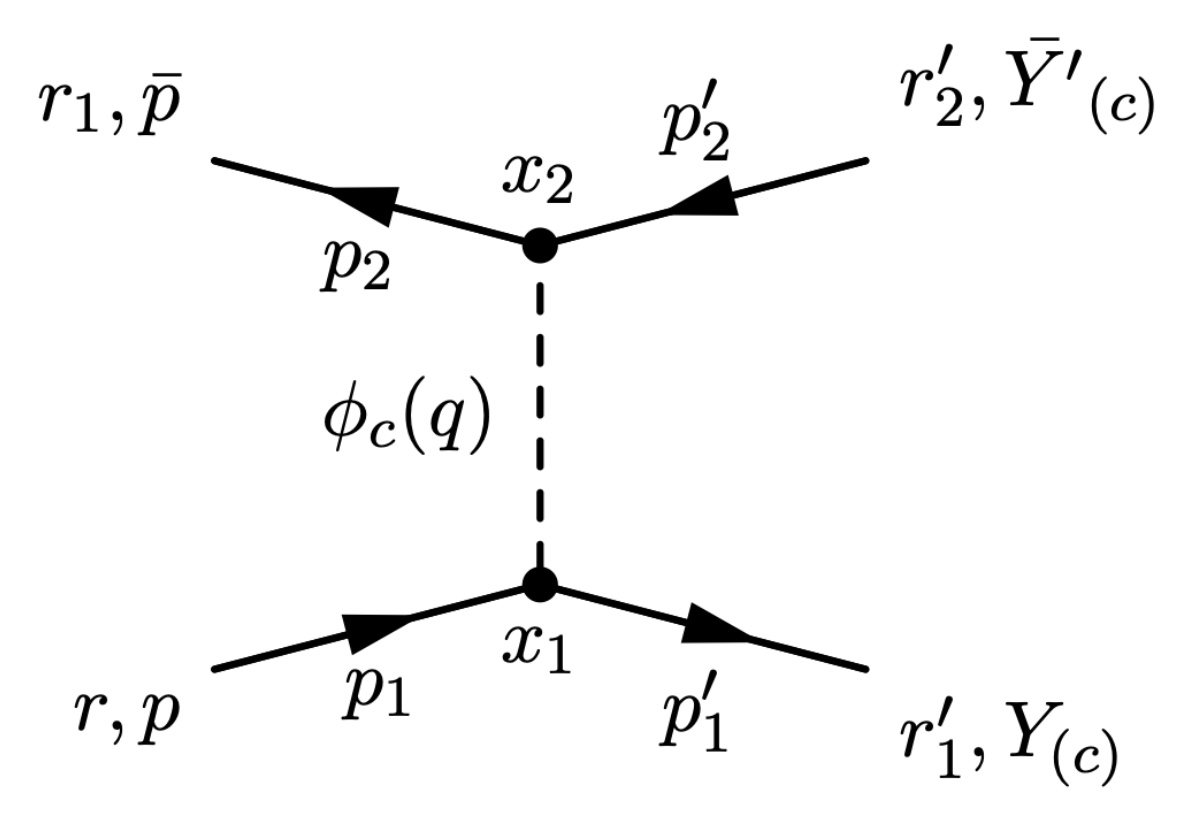}
\caption{ Tree-level diagram for the reactions $p\bar p \to Y_{c}\bar Y_{c}'$, where $Y_{c}$ and $\Bar{Y}'_{c}$ represent charmed baryons. $\phi_{c}$, in the intermediate line, represents the pseudoscalar and vector $D$ mesons, respectively.}
    \label{Feynman diagram}
\end{figure}
%\vspace{-3mm}
In this section, we aim to calculate the differential cross-sections of %the proton-antiproton scattering from $p\bar p \to Y_{c}\bar Y_{c}'$ such as
\( p\bar{p} \rightarrow \Lambda_{c}\bar{\Lambda}_{c}, \Sigma_{c}\bar{\Sigma}_{c},\) \( \Sigma_{c}^{*}\bar{\Sigma}_{c}^{*}, 
\Sigma_{c}^{*}\bar{\Sigma}_{c},\Sigma_{c}\bar{\Sigma}_{c}^{*}, \Sigma_{c}\bar{\Lambda}_{c}, \Lambda_{c}\bar{\Sigma}_{c}, \Sigma_{c}^{*}\bar{\Lambda}_{c}, \Lambda_{c}\bar{\Sigma}_{c}^{*} \)  processes. 
For those processes, we will investigate the consequences of the HQSS and its violation effects in the charmed baryon productions and the Feynman diagram of these processes have been shown in \cref{Feynman diagram}. %where the exchanged particles, $\phi_{c}$, refer to pseudoscalar and vector $D$ mesons respectively.

\subsection{The scattering amplitudes}
According to the $SU(2)_f$ effective Lagrangians in terms of \emph{conserving} and \emph{violating} parts, the scattering amplitudes of charmed productions with the exchanges of the pseudoscalar and vector $D$ mesons are written as follows,\\ \\
\textbf{Process: \,\,$p\bar p\to \Lambda_c\bar \Lambda_c$}
\begin{align}
\begin{split}
\mathcal{M}_{\mathcal{C}}\,\,=\,\,\,\,\,\,&g_{2}^2\,\Gamma_{N(P)}\,G(q)\,\Gamma_{\Bar{N}(P)}
+\frac{g_{4}^{2}}{m_{D}^{2}}\,\Gamma_{N(A)}\,G(q)\,\,\Gamma_{\bar{N}(A)}
\\
-&\frac{g_{2}\,g_{4}}{m_{D}}\,i\Gamma_{N(P)}\,G(q)\,\Gamma_{\Bar{N}(A)}+\frac{g_{4}\,g_{2}}{m_{D}}\,i\Gamma_{N(A)}\,G(q)\,\Gamma_{\Bar{N}(P)}\,,
\\ 
\mathcal{M}_{\mathcal{V}}\,\,=\,\,\,\,\,\,&f_{2}^2\,\Gamma_{N(V)}^\mu\,G_{\mu\nu}(q)\,\Gamma_{\bar{N}(V)}^\nu +\frac{h_{2}^{2}}{m_{D^{*}}^{2}}\,\Gamma_{N(T)}^\mu\,G_{\mu\nu}(q)\,\,\Gamma_{\bar{N}(T)}^\nu\\
+&\frac{f_{2}\,h_{2}}{m_{D^{*}}}\,i\Gamma_{N(V)}^\mu\,G_{\mu\nu}(q)\,\Gamma_{\bar{N}(T)}^\nu  -\frac{h_{2}\,f_{2}}{m_{D^{*}}}\,i\Gamma_{N(T)}^\mu\,G_{\mu\nu}(q)\,\Gamma_{\bar{N}(V)}^\nu\,.
\end{split}
\end{align}
\textbf{Process: \,\,$p\bar p\to \Sigma_c\bar \Sigma_c$}
\begin{align}
\begin{split}
\mathcal{M}_{\mathcal{C}}\,\,=\,\,\,\,\,\,&f_{1}^{2}\,\Gamma_{N(V)}^\mu\,G_{\mu\nu}(q)\,\Gamma_{\bar{N}(V)}^\nu\,,
\\
\mathcal{M}_{\mathcal{V}}\,\,=\,\,\,\,\,\,&g_{1}^2\,\Gamma_{N(P)}\,G(q)\,\Gamma_{\Bar{N}(P)}+
\frac{g_{3}^{2}}{m_{D}^{2}}\,\Gamma_{N(A)}\,G(q)\,\,\Gamma_{\bar{N}(A)}\\-&\frac{g_{1}\,g_{3}}{m_{D}}\,i\Gamma_{N(P)}\,G(q)\,\Gamma_{\Bar{N}(A)}+\frac{g_{3}\,g_{1}}{m_{D}}\,i\Gamma_{N(A)}\,G(q)\,\Gamma_{\Bar{N}(P)}\\+&\frac{h_{1}^{2}}{m_{D^{*}}^{2}}\,\Gamma_{N(T)}^\mu\,G_{\mu\nu}(q)\,\,\Gamma_{\bar{N}(T)}^\nu\,.
\end{split}
\end{align}
\textbf{Process: \,\,$p\bar{p}\rightarrow \Sigma_{c}^{*}\bar{\Sigma}_{c}^{*}$}
\begin{align}
\begin{split}
\mathcal{M}_{\mathcal{C}}\,\,=\,\,\,\,\,\,&f_{3}^2\,\Gamma_{N(V^{*})}^\mu\,G_{\mu\nu}(q)\,\Gamma_{\bar{N}(V^{*})}^\nu -\frac{h_{3}^{2}}{4m_{D^{*}}^{2}}\,\Gamma_{N(T)}^\mu\,G_{\mu\nu}(q)\,\,\Gamma_{\bar{N}(T)}^\nu\\-&
\frac{f_{3}\,h_{3}}{2m_{D^{*}}}\,\Gamma_{N(V^{*})}^\mu\,G_{\mu\nu}(q)\,\Gamma_{\bar{N}(T^{*})}^\nu+\frac{h_{3}\,f_{3}}{2m_{D^{*}}}\Gamma_{N(T^{*})}^\mu\,G_{\mu\nu}(q)\,\Gamma_{\bar{N}(V^{*})}^\nu\,,
\\
\mathcal{M}_{\mathcal{V}}\,\,=\,\,\,\,\,\,&\frac{h_{4}^{2}}{m_{D^{*}}^{2}}\,\Gamma_{N(T^{*})}^\mu\,G_{\mu\nu}(q)\,\,\Gamma_{\bar{N}(T^{*})}^\nu\,.
\end{split}
\end{align}
\textbf{Process: $p\bar{p}\rightarrow \Sigma_{c}^{*}\bar{\Sigma}_{c}$}
\begin{align}
\begin{split}
\mathcal{M}_{\mathcal{C}}\,\,=\,\,\,\,\,\,&\frac{f_{1}f_{3}}{2}\,\Gamma_{N(V)}^\mu\,G_{\mu\nu}\,(q)\,\Gamma_{\Bar{N}(V^{*})}^\nu +\frac{f_{3}f_{1}}{2}\,\Gamma_{N(V^{*})}^\mu\,G_{\mu\nu}\,(q)\,\Gamma_{\Bar{N}(V)}^\nu
\\+&\frac{f_{1}\,h_{3}}{4m_{D^{*}}}\,\Gamma_{N(V)}^\mu\,G_{\mu\nu}(q)\,\Gamma_{\bar{N}(A^{*})}^\nu +\frac{h_{3}\,f_{1}}{4m_{D^{*}}}\,\Gamma_{N(A^{*})}^\mu\,G_{\mu\nu}(q)\,\Gamma_{\bar{N}(V)}^\nu\,,
\\
\mathcal{M}_{\mathcal{V}}\,\,=-&\frac{h_{1}h_{4}}{2m_{D^{*}}^{2}}\,\Gamma_{\Bar{N}(T)}^\mu\,G_{\mu\nu}(q)\,\,\Gamma_{N(T^{*})}^\nu -\frac{h_{4}h_{1}}{2m_{D^{*}}^{2}}\,\Gamma_{N(T^{*})}^\mu\,G_{\mu\nu}(q)\,\,\Gamma_{\bar{N}(T)}^\nu\,.
\end{split}
\end{align}
\textbf{Process: $p\bar{p}\rightarrow \Sigma_{c}\bar{\Sigma}_{c}^{*}$}
\begin{align}
\begin{split}
\mathcal{M}_{\mathcal{C}}\,\,=\,\,\,\,\,\,&\frac{f_{1}f_{3}}{2}\,\Gamma_{N(V)}^\mu\,G_{\mu\nu}\,(q)\,\Gamma_{\Bar{N}(V^{*})}^\nu +\frac{f_{3}f_{1}}{2}\,\Gamma_{N(V^{*})}^\mu\,G_{\mu\nu}\,(q)\,\Gamma_{\Bar{N}(V)}^\nu\,
\\-&\frac{f_{1}\,h_{3}}{4m_{D^{*}}}\,\Gamma_{N(V)}^\mu\,G_{\mu\nu}(q)\,\Gamma_{\bar{N}(A^{*})}^\nu -\frac{h_{3}\,f_{1}}{4m_{D^{*}}}\,\Gamma_{N(A^{*})}^\mu\,G_{\mu\nu}(q)\,\Gamma_{\bar{N}(V)}^\nu\,,
\\
\mathcal{M}_{\mathcal{V}}\,\,=-&\frac{h_{1}h_{4}}{2m_{D^{*}}^{2}}\,\Gamma_{N(T)}^\mu\,G_{\mu\nu}(q)\,\,\Gamma_{\bar{N}(T^{*})}^\nu -\frac{h_{4}h_{1}}{2m_{D^{*}}^{2}}\,\Gamma_{N(T^{*})}^\mu\,G_{\mu\nu}(q)\,\,\Gamma_{\bar{N}(T)}^\nu\,.
\end{split}
\end{align}
\textbf{Process: $p\bar{p}\rightarrow \Lambda_{c}\bar{\Sigma}_{c}$}
\begin{align}
\begin{split}
\mathcal{M}_{\mathcal{V}}\,\,=\,\,\,\,\,\,&\frac{f_{2}h_{1}}{2m_{D^{*}}}\,i\Gamma_{N(V)}^\mu\,G_{\mu\nu}(q)\,\,\Gamma_{\bar{N}(T)}^\nu +\frac{h_{1}f_{2}}{2m_{D^{*}}}\,i\Gamma_{\Bar{N}(T)}^\mu\,G_{\mu\nu}(q)\,\,\Gamma_{\bar{N}(V)}^\nu
\\+&\frac{h_{1}h_{2}}{2m_{D^{*}}^{2}}\,\Gamma_{\Bar{N}(T)}^\mu\,G_{\mu\nu}(q)\,\,\Gamma_{N(T)}^\nu +\frac{h_{2}h_{1}}{2m_{D^{*}}^{2}}\,\Gamma_{N(T)}^\mu\,G_{\mu\nu}(q)\,\,\Gamma_{\bar{N}(T)}^\nu\,.
\end{split}
\end{align}
\textbf{Process: $p\bar{p}\rightarrow \Sigma_{c}\bar{\Lambda}_{c}$}
\begin{align}
\begin{split}
\mathcal{M}_{\mathcal{V}}\,\,=-&\frac{f_{2}h_{1}}{2m_{D^{*}}}\,i\Gamma_{\bar{N}(V)}^\mu\,G_{\mu\nu}(q)\,\,\Gamma_{\bar{N}(T)}^\nu +\frac{h_{1}f_{2}}{2m_{D^{*}}}\,i\Gamma_{N(T)}^\mu\,G_{\mu\nu}(q)\,\,\Gamma_{\Bar{N}(V)}^\nu
\\
\,\,\,\,\,+&\frac{h_{1}h_{2}}{2m_{D^{*}}^{2}}\,\Gamma_{N(T)}^\mu\,G_{\mu\nu}(q)\,\,\Gamma_{\bar{N}(T)}^\nu +\frac{h_{2}h_{1}}{2m_{D^{*}}^{2}}\,\Gamma_{\Bar{N}(T)}^\mu\,G_{\mu\nu}(q)\,\,\Gamma_{N(T)}^\nu\,.
\end{split}
\end{align}
\textbf{Process: $p\bar{p}\rightarrow \Lambda_{c}\bar{\Sigma}_{c}^{*}$}
\begin{align}
\begin{split}
\mathcal{M}_{\mathcal{V}}\,=-&\frac{h_{4}f_{2}}{2m_{D^{*}}}\,i\Gamma_{\Bar{N}(T^{*})}^\mu\,G_{\mu\nu}(q)\,\Gamma_{N(V)}^\nu
-\frac{f_{2}h_{4}}{2m_{D^{*}}}\,i\Gamma_{N(V)}^\mu\,G_{\mu\nu}(q)\,\Gamma_{\bar{N}(T^{*})}^\nu
\\-&\frac{h_{2}h_{4}}{2m_{D^{*}}^{2}}\,\Gamma_{N(T)}^\mu\,G_{\mu\nu}(q)\,\Gamma_{\bar{N}(T^{*})}^\nu -\frac{h_{4}h_{2}}{2m_{D^{*}}^{2}}\,\Gamma_{\Bar{N}(T^{*})}^\mu\,G_{\mu\nu}(q)\,\Gamma_{\bar{N}(T)}^\nu\,.
\end{split}
\end{align}
\textbf{Process: $p\bar{p}\rightarrow \Sigma_{c}^{*}\bar{\Lambda}_{c}$}
\begin{align}
\begin{split}
\mathcal{M}_{\mathcal{V}}\,\,=\,\,\,\,\,&\frac{h_{4}f_{2}}{2m_{D^{*}}}\,i\Gamma_{N(T^{*})}^\mu\,G_{\mu\nu}(q)\,\Gamma_{\bar{N}(V)}^\nu
+\frac{f_{2}h_{4}}{2m_{D^{*}}}\,i\Gamma_{\Bar{N}(V)}^\mu\,G_{\mu\nu}(q)\,\Gamma_{N(T^{*})}^\nu
\\-&\frac{h_{2}h_{4}}{2m_{D^{*}}^{2}}\,\Gamma_{N(T^{*})}^\mu\,G_{\mu\nu}(q)\,\Gamma_{\bar{N}(T)}^\nu -\frac{h_{4}h_{2}}{2m_{D^{*}}^{2}}\,\Gamma_{\Bar{N}(T)}^\mu\,G_{\mu\nu}(q)\,\Gamma_{N(T^{*})}^\nu\,.
\end{split}
\end{align}
Here, $\mathcal{M}_\mathcal{C}$ and $\mathcal{M}_\mathcal{V}$ denote the \emph{conserving} and \emph{violating} scattering amplitudes, respectively. Note that, due to HQSS, the charmed baryons $\Sigma_{c}$ and $\Sigma_{c}^{*}$ are spin partners, therefore conserving HQSS amplitudes do not appear in the $p\bar{p}\rightarrow\Lambda_{c}\Bar{\Sigma}_{c}$, $\Sigma_{c}\Bar{\Lambda}_{c}$, $\Lambda_{c}\Bar{\Sigma}^{*}_{c}$, and $\Sigma_{c}^{*}\Bar{\Lambda}_{c}$ the processes. In addition, the $\Gamma$ notations appear in the scattering amplitudes above and they are defined by, 
\begin{align}
\begin{split}
&\Gamma_{N(P)} = \,\Bar{u}_{Y_c}\,i\,\gamma_{5}\,u_{N}, \quad \Gamma_{\Bar{N}(P)} = \,-\Bar{v}_{\Bar{N}}\,i\,\gamma_{5}\,v_{\Bar{Y}_c'}\,,
\\
&\Gamma_{N(A)} = \,\Bar{u}_{Y_c}\,\gamma^{\mu}\,\gamma_{5}\,q_{\mu}\,u_{N}, \quad \Gamma_{\Bar{N}(A)} = \,-\Bar{v}_{\Bar{N}}\,\gamma^{\nu}\,\gamma_{5}\,q_{\nu}\,v_{\Bar{Y}_c'}\,,
\\
&\Gamma_{N(A^{*})}^\mu = \,\Bar{u}^{\mu}_{Y_c}\,\gamma^{\alpha}\,\gamma_{5}\,q_{\alpha}\,u_{N}, \quad \Gamma_{\Bar{N}(A^{*})}^\nu = \,-\Bar{v}_{\Bar{N}}\,\gamma^{\beta}\,\gamma_{5}\,q_{\beta}\,v_{\Bar{Y}_c'}^{\nu}\,,
\\
&\Gamma_{N(V)}^\mu = \,\Bar{u}_{Y_c}\,\gamma^{\mu}\,u_{N}, \quad \Gamma_{\Bar{N}(V)}^\nu = \,-\Bar{v}_{\Bar{N}}\,\gamma^{\nu}\,v_{\Bar{Y}_c'}\,,
\\
&\Gamma_{\Bar{N}(V^{*})}^\mu = \,\Bar{u}^{\mu}_{Y_c}\,\gamma_{5}\,u_{N}, \quad \Gamma_{N(V^{*})}^\nu =-\Bar{v}_{\Bar{N}}\,\gamma_{5}\,v_{\Bar{Y}_c'}^{\nu}\,, 
\\
&\Gamma_{N(T)}^{\nu} = \,\Bar{u}_{Y_c}\,\sigma^{\mu\nu}q_{\mu}\,u_{N}, \quad \Gamma_{\Bar{N}(T)}^{\beta} = \,-\Bar{v}_{\Bar{N}}\,\sigma^{\alpha\beta}q_{\alpha}\,v_{\Bar{Y}_c'}\,,
\\
&\Gamma_{N(T^{*})}^{\beta} = \,\epsilon^{\mu\nu\alpha\beta}\Bar{u}_{Y_c,\,\mu}\,\gamma_{\nu}q_{\alpha}\,u_{N},\quad \Gamma_{\Bar{N}(T^{*})}^{\kappa} = \,-\epsilon^{\rho\sigma\eta\kappa}\Bar{v}_{\Bar{N}}\, \gamma_{\sigma}q_{\eta}\,v_{\Bar{Y}_c',\,\rho}\,,
\end{split}
\end{align} 
where $\Gamma_{N(V^{*})}$, $\Gamma_{N(A^{*})}$ and $\Gamma_{N(T^{*})}$ represent for charmed baryon spin 3/2 vertices and the Feynman propagators for pseudoscalar $D$ meson spin-0 and vector $D$ meson  spin-1 are defined by 
\begin{align}
G(q) = \frac{i}{q-m^{2}_{D}}, \quad G_{\mu\nu}(q) = \frac{i}{q-m^{2}_{D^{*}}}\left(-g_{\mu\nu}+\frac{q_{\mu}q_{\nu}}{m^{2}_{D^*}}\right).
\end{align}
\subsection{Differential Cross-Sections}
It is well known that a study of the scattering of the composite particles, precisely hadrons needs the form factor to regulate the amplitudes. % the overestimation of the cross-sections and the infinite size of hadrons at higher energies, 
In this work, we include the phenomenological form factors at the vertices, adopted from Ref.~\cite{Sangkhakrit:2020wyi}, which are fixed by comparison with experimental data on the production of strangeness and then use them to predict the charmed baryon productions. 
The form factors used in this work are given by
\begin{align} 
F(t) & =a^2 \frac{\Lambda^4}{\Lambda^4+\left(t-m_\phi^2\right)^2} \,,
\\ F_n(t) & =a\left(\frac{\Lambda^2}{\Lambda^2-t}\right)^n, \quad(n=1,2)
\label{formfactor}
\end{align}
where the form factor \( F(t) \) has parameters \( a = 0.46 \) and \( \Lambda = 0.63 \, \text{GeV} \). For \( F_1(t) \), the parameters are \( a = 0.285 \) and \( \Lambda = 0.7 \, \text{GeV} \), while for \( F_2(t) \), the parameters are \( a = 0.285 \) and \( \Lambda = 0.99 \, \text{GeV} \). The total amplitudes of the reactions, $p\bar p \to Y_c\bar Y_c$ are written as
\begin{align}
\mathcal{M}_{p \bar{p} \rightarrow Y_{c} \bar{Y}_{c}^{\prime}}=\left\{\begin{array}{l}\mathcal{M}_D F_D+\mathcal{M}_{D^*} F_{D^*}, \\ \mathcal{M}_D F_{n, D}^2+\mathcal{M}_{D^*} F_{n, D^*}^2.\end{array}\right.
\end{align}
The form factors \( F_{(D,D^{*})} \) and \( F_{n(D,D^{*})} \) are already mentioned in Eq.~(\ref{formfactor}) for the pseudoscalar $D$ meson ($D$) and the vector $D$ meson ($D^{*}$). 
It is common practice to use various functional forms and cutoff values for $t$-channel form factors \cite{Haidenbauer:2014rva,Shyam:2016uxa,Shyam:2017gqp}.
The differential cross-sections as a function of $t$ is calculated from
\begin{align}
    \frac{d\sigma}{dt} &= \frac{1}{64\pi(p_{cm})^2s} \left\langle |\mathcal{M}|^2 \right\rangle. 
\end{align}
Here, \( p_{\text{cm}} \) is the relative momentum of \( p \) and \( \bar{p} \) in the center-of-mass frame, \( s \) is the Mandelstam variable. The term \( \left\langle |\mathcal{M}|^2 \right\rangle \) is the spin-averaged and summed amplitude, given by
\begin{align}
\left\langle |\mathcal{M}|^2 \right\rangle = \frac{1}{4} \sum_{s_3, s_4} |\mathcal{M}|^2.
\label{spinsum}
\end{align}
where the sum runs over the spins of the final-state particles. After averaging and summing over spins Eq.~(\ref{spinsum}) the interference term vanishes. This is because the spin averaging eliminates interference between spin configurations. 
The trace analysis reveals that the interference term would mix $D$ and $D^*$ mesons, which is prohibited by parity conservation. 
In this study, the differential cross-sections, $d\sigma/dt$, in section \ref{sec-4} are presented as a function of $t_{max}-t$. 
For a specific energy value, \( t \) varies from \( t_{\text{min}} \) to \( t_{\text{max}} \) (i.e., \( t_{\text{max}} - t \) varies from 0 to \( t_{\text{max}} - t_{\text{min}} \)). One can write the explicit form of the $t_{\max }^{\min }$ as,
\begin{align}
\begin{split}
t_{\max }^{\min } & = m_N^2+m_{Y_{c}}^2-\frac{1}{2 s}\Big[s(s+m_{Y_{c}}^2-m_{Y_{c}^{\prime}}^2) \pm \sqrt{s(s-4 m_N^2)(s-(m_{Y_{c}}+m_{Y_{c}^{\prime}})^2)}\\&\times \sqrt{(s-(m_{Y_{c}}-m_{Y_{c}^{\prime}})^2)}\Big]\,.
\end{split}
\end{align}
%%%%%%%%%%%%%%%%%%%%%%%%%%%%%%%%%%%%%%%%%%%%%%%%%%%%%%%%%%%%%%%%%%%%%%%%%%%%%%%%%%%%%%%%%%%%%%%%%%%%%%%%%%%%%%%%%%%%%%%%%%%%%%%%%%%%%%%%%%%%%%
\section{Results and Discussions}\label{sec-4}

In this section, the numerical results of the differential cross-sections, \(d\sigma/dt\), are presented as a function of \(t - t_{\text{max}}\) for the reaction \(p\bar{p} \to Y_{c}\bar{Y}_{c}'\), evaluated for each form factor at \(p_{\text{Lab}} = 15\, \text{GeV/c}\) in figs.~\ref{figLcLc_combined} to \ref{fig:ScsLc_combined}. 
The green bands represent the \emph{conserving} production rates, the red bands show the \emph{violating} parts, and the purple bands illustrate the total contributions.
\vspace{-3mm}
\subsection{The scattering amplitudes of the $p\bar p\to Y_c\bar Y^{'}_c$ reactions}
\label{The scattering amplitudes}
\vspace{-3mm}
\begin{figure}[h]
    \centering
    \begin{subfigure}{0.328\textwidth}
        \centering
        \includegraphics[width=\linewidth]{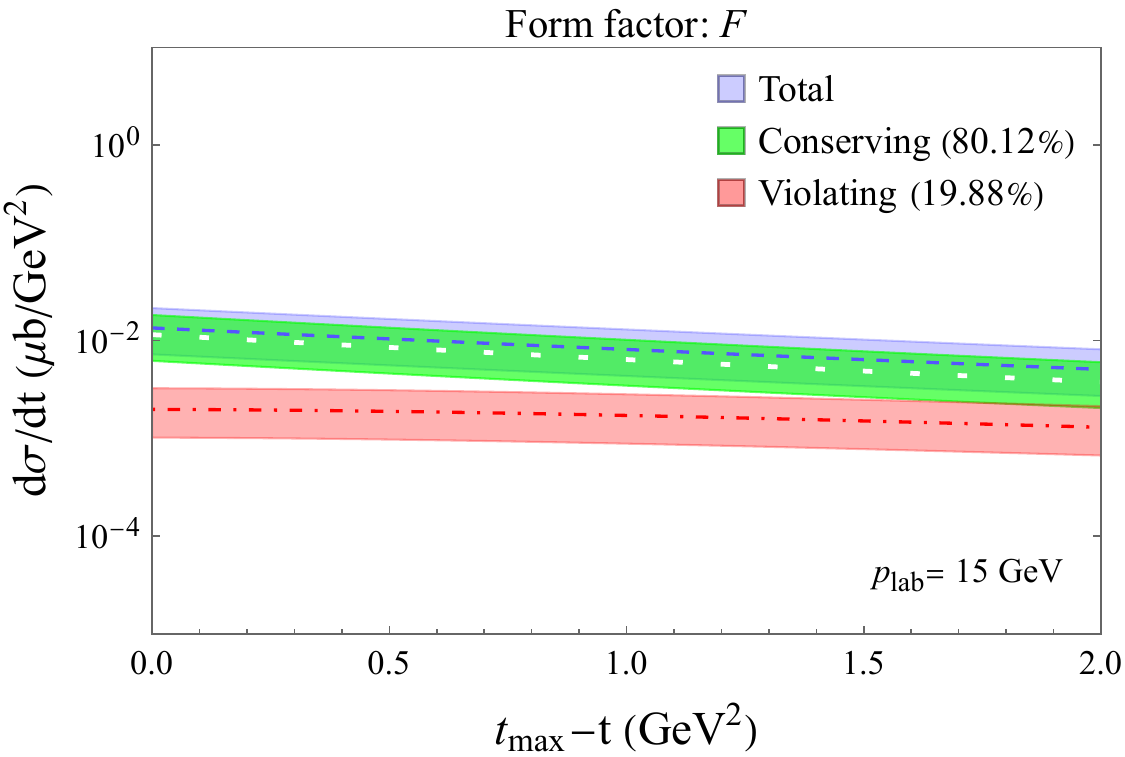}
        \caption{}
        \label{fig:LcLc CF0}
    \end{subfigure}
    \hfill
    \begin{subfigure}{0.328\textwidth}
        \centering
        \includegraphics[width=\linewidth]{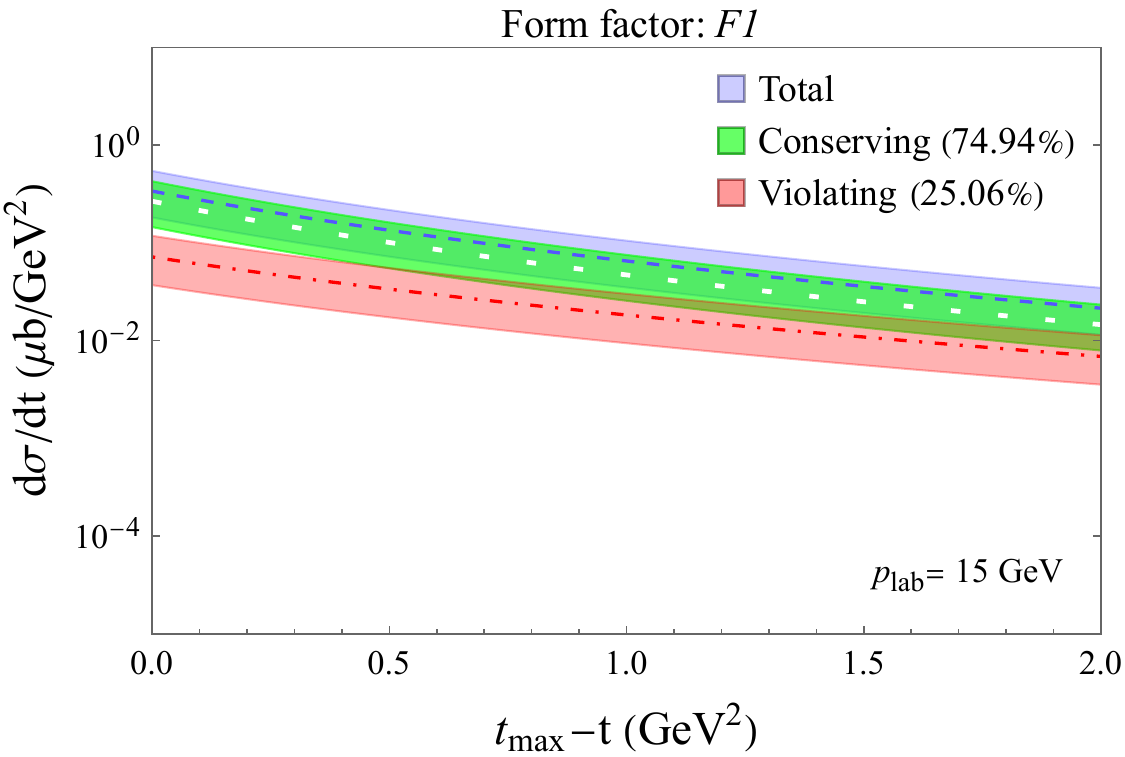}
        \caption{}
        \label{fig:LcLc CF1}
    \end{subfigure}
    \hfill
    \begin{subfigure}{0.328\textwidth}
        \centering
        \includegraphics[width=\linewidth]{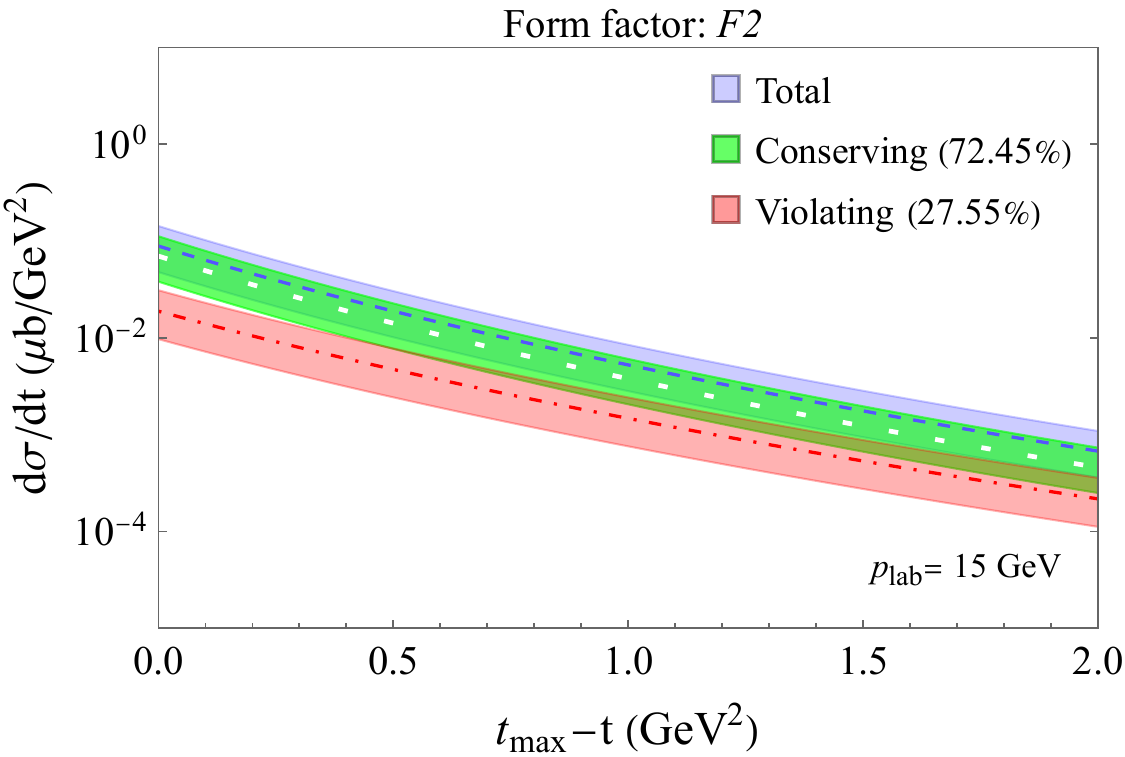}
        \caption{}
        \label{fig:LcLc CF2}
    \end{subfigure}
    \caption{Differential cross-section for $p\bar p\to \Lambda_c\bar \Lambda_c$ in term of conserving and violating contributions at 15 GeV.}
    \label{figLcLc_combined}
\end{figure}
\FloatBarrier
\vspace{-3mm}
\begin{figure}[h]
    \centering
    \begin{subfigure}{0.328\textwidth}
        \centering
        \includegraphics[width=\linewidth]{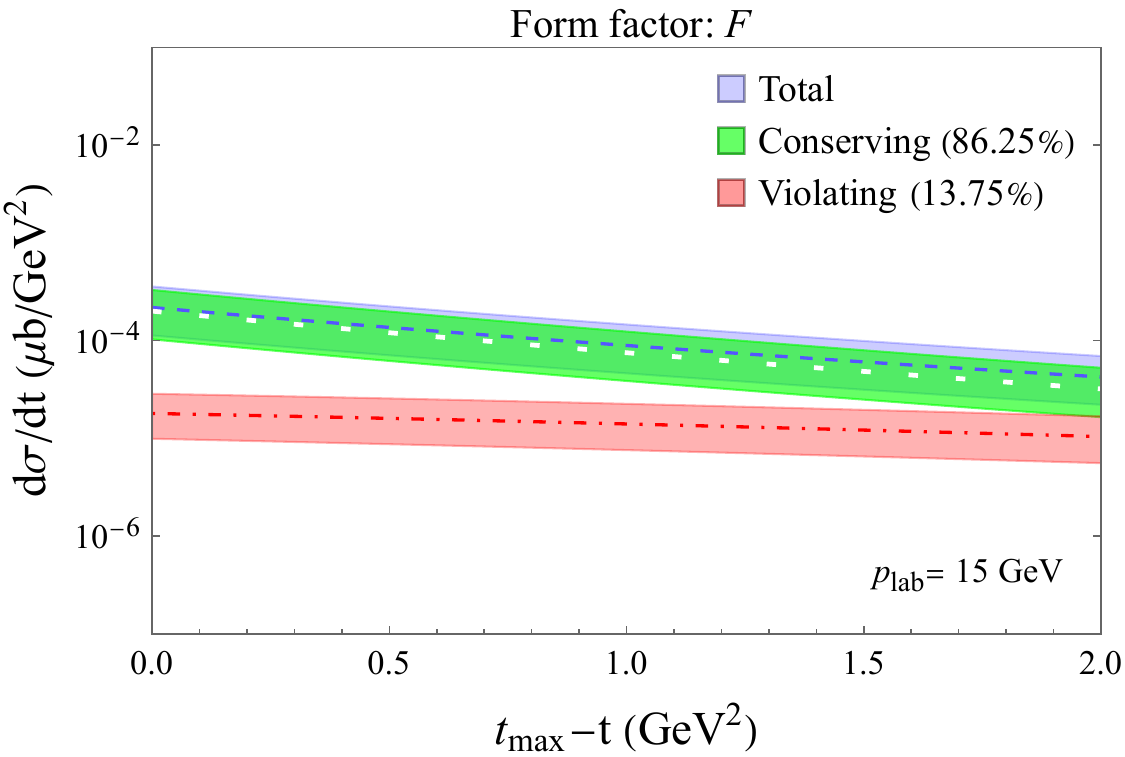}
        \caption{}
        \label{fig:ScSc F0}
    \end{subfigure}
    \hfill
    \begin{subfigure}{0.328\textwidth}
        \centering
        \includegraphics[width=\linewidth]{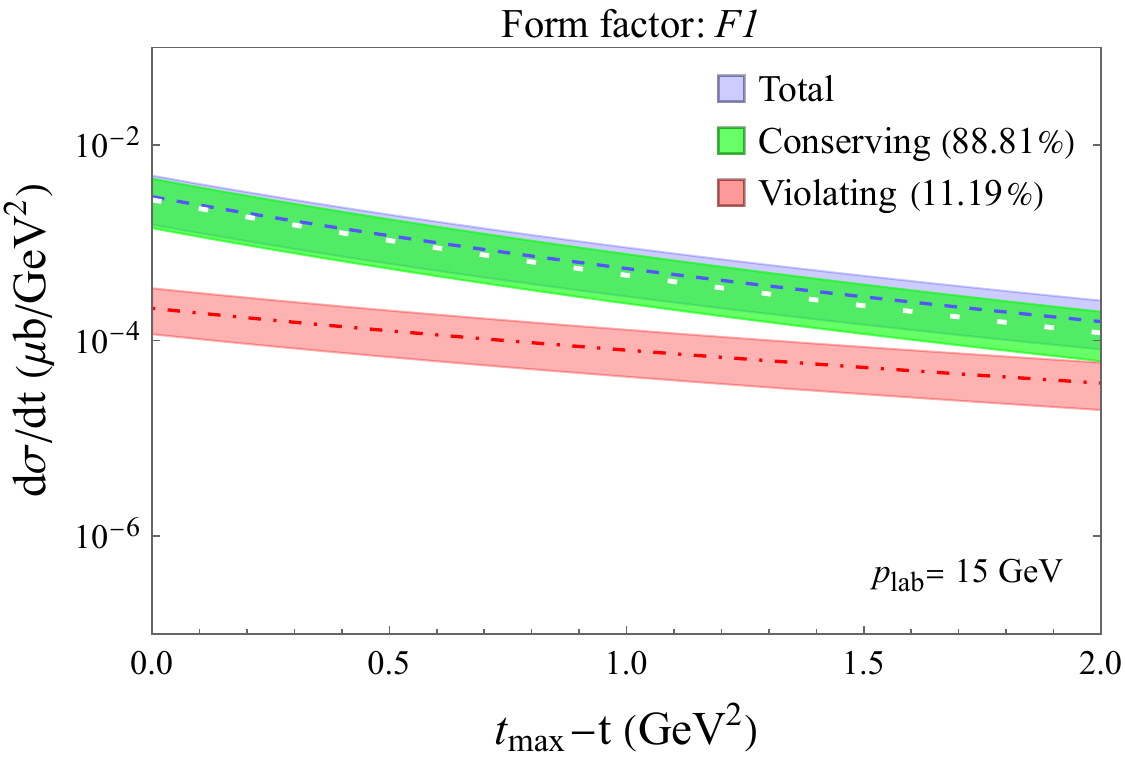}
        \caption{}
        \label{fig:ScSc F1}
    \end{subfigure}
    \hfill
    \begin{subfigure}{0.328\textwidth}
        \centering
        \includegraphics[width=\linewidth]{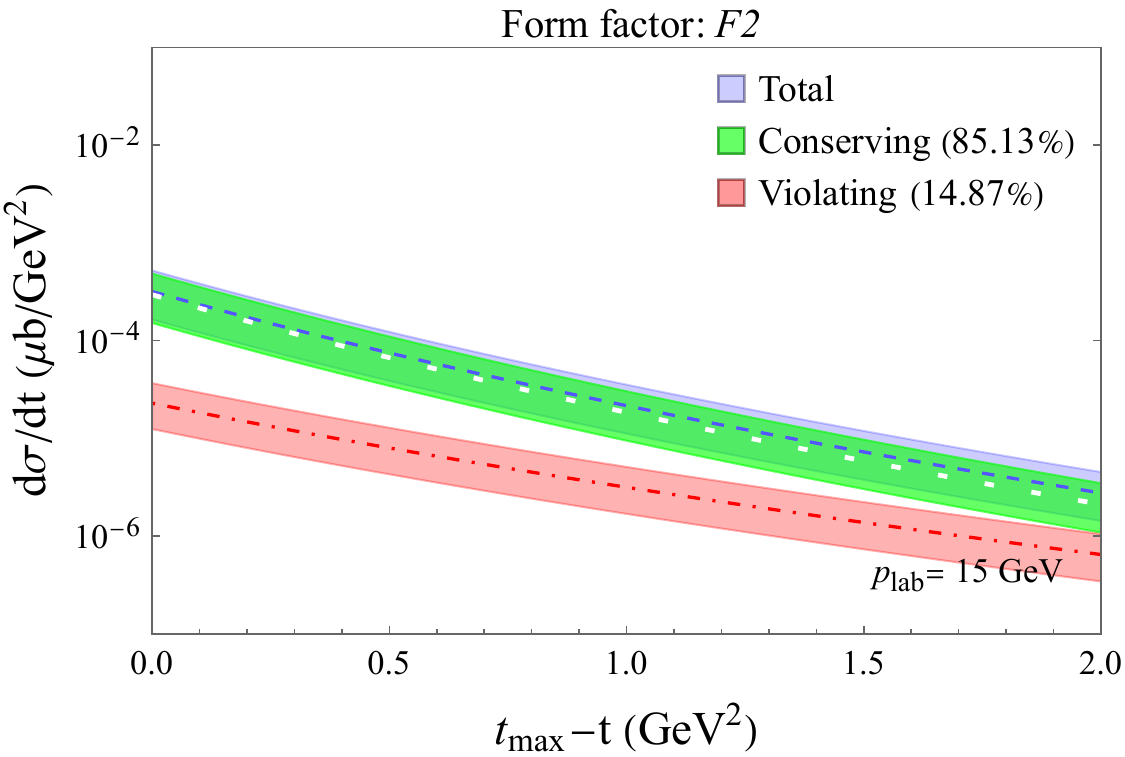}
        \caption{}
        \label{fig:ScSc F2}
    \end{subfigure}
\caption{Differential cross-section for $p\bar p\to \Sigma_c\bar \Sigma_c$ in term of conserving and violating contributions at 15 GeV.}
    \label{fig:ScSc_combined}
\end{figure}
\begin{figure}[h]
    \centering
    \begin{subfigure}{0.328\textwidth}
        \centering
        \includegraphics[width=\linewidth]{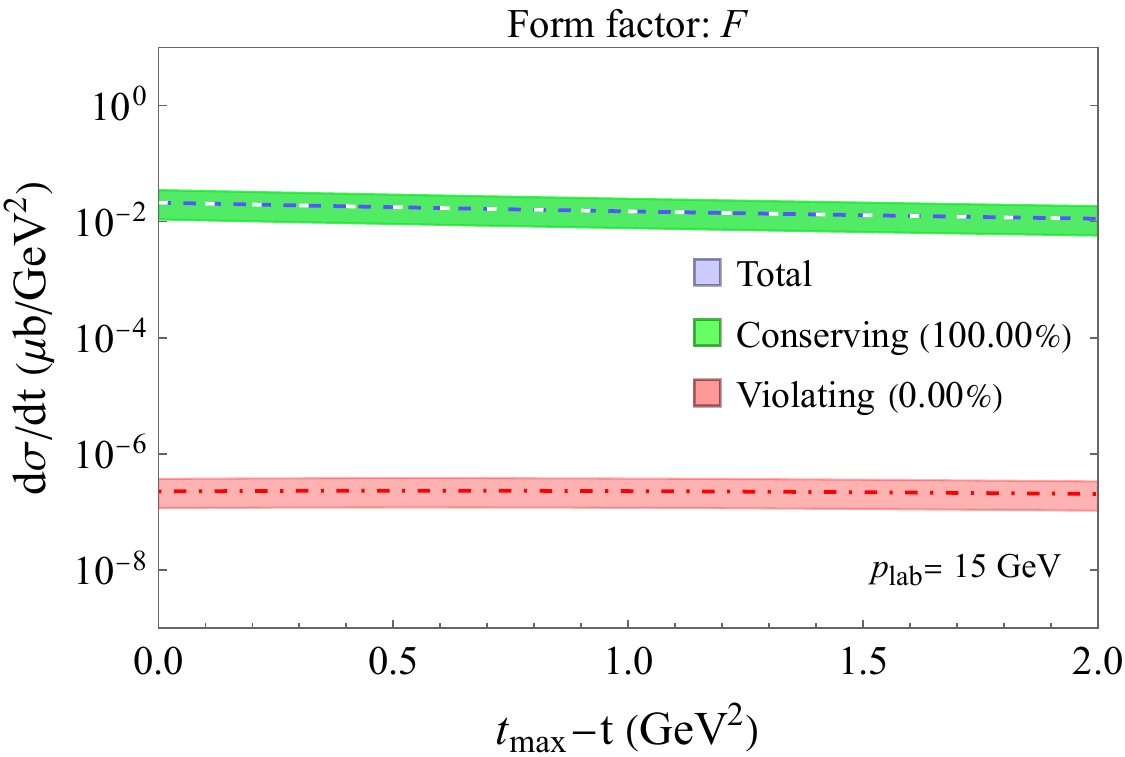}
        \caption{}
        \label{fig:ScsScs F0}
    \end{subfigure}
    \hfill
    \begin{subfigure}{0.328\textwidth}
        \centering
        \includegraphics[width=\linewidth]{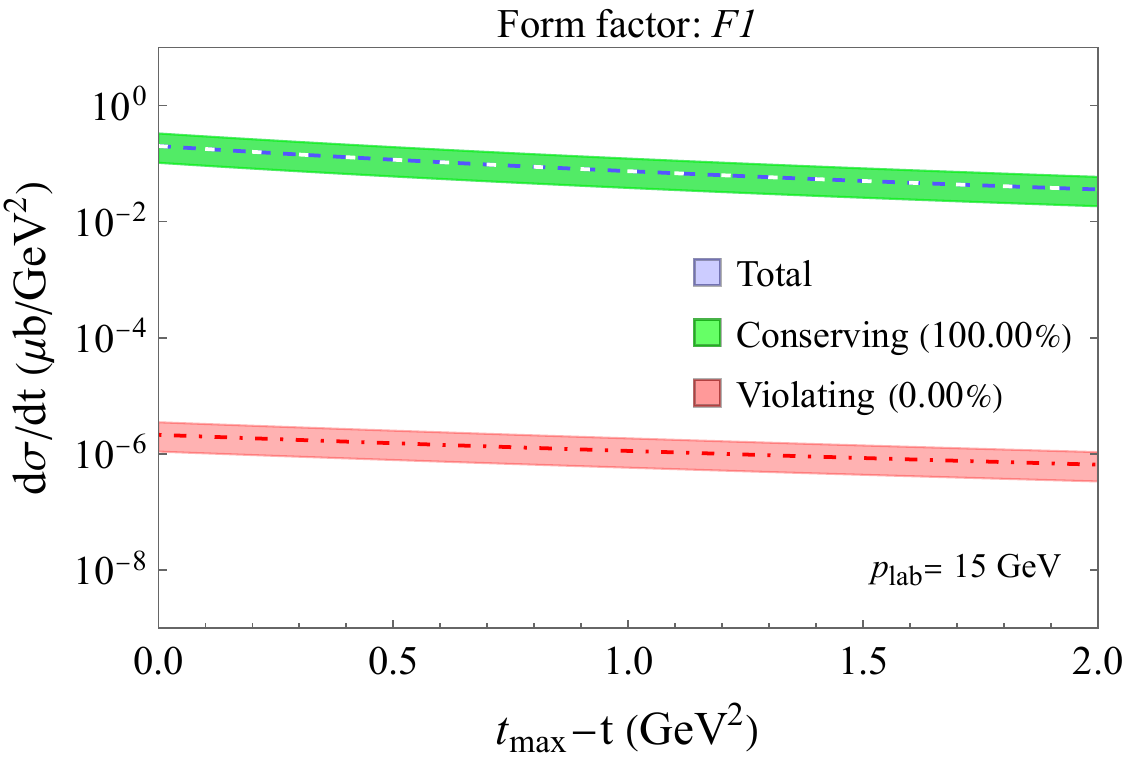}
        \caption{}
        \label{fig:ScsScs F1}
    \end{subfigure}
    \hfill
    \begin{subfigure}{0.328\textwidth}
        \centering
        \includegraphics[width=\linewidth]{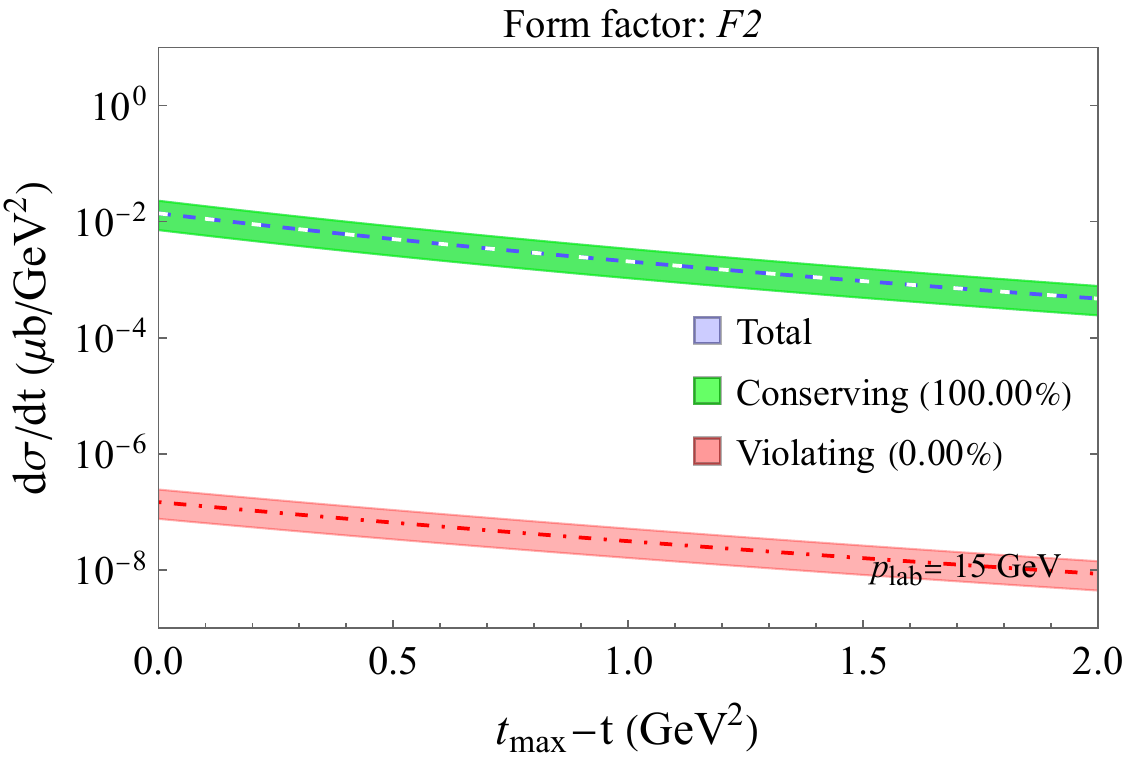}
        \caption{}
        \label{fig:ScsScs F2}
    \end{subfigure}
\caption{Differential cross-section for $p\bar p\to\Sigma_c^*\bar \Sigma_c^*$ in term of conserving and violating contributions at 15 GeV. Note that the purple band completely overlaps with the green band, showing the dominance of the HQSS-conserving contribution.}
    \label{fig:ScsScs_combined}
\end{figure}
\begin{figure}[h]
    \centering
    \begin{subfigure}{0.328\textwidth}
        \centering
        \includegraphics[width=\linewidth]{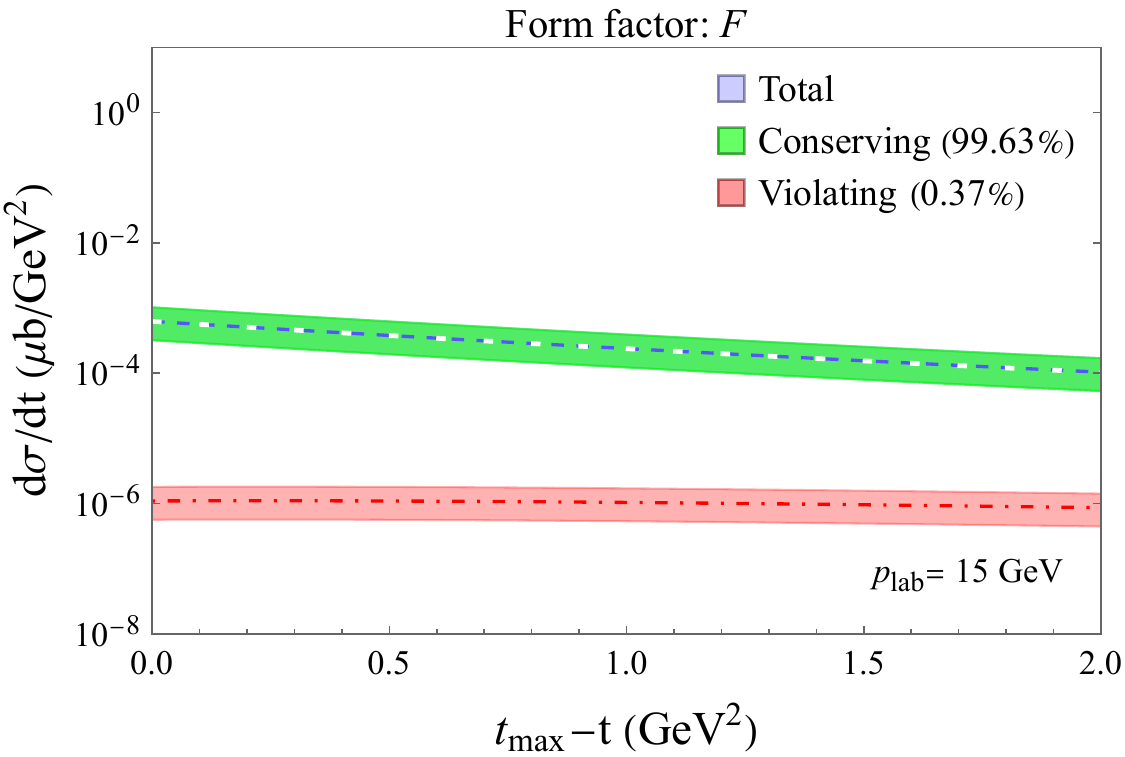}
        \caption{}
        \label{fig:ScsSc F0}
    \end{subfigure}
    \hfill
    \begin{subfigure}{0.328\textwidth}
        \centering
        \includegraphics[width=\linewidth]{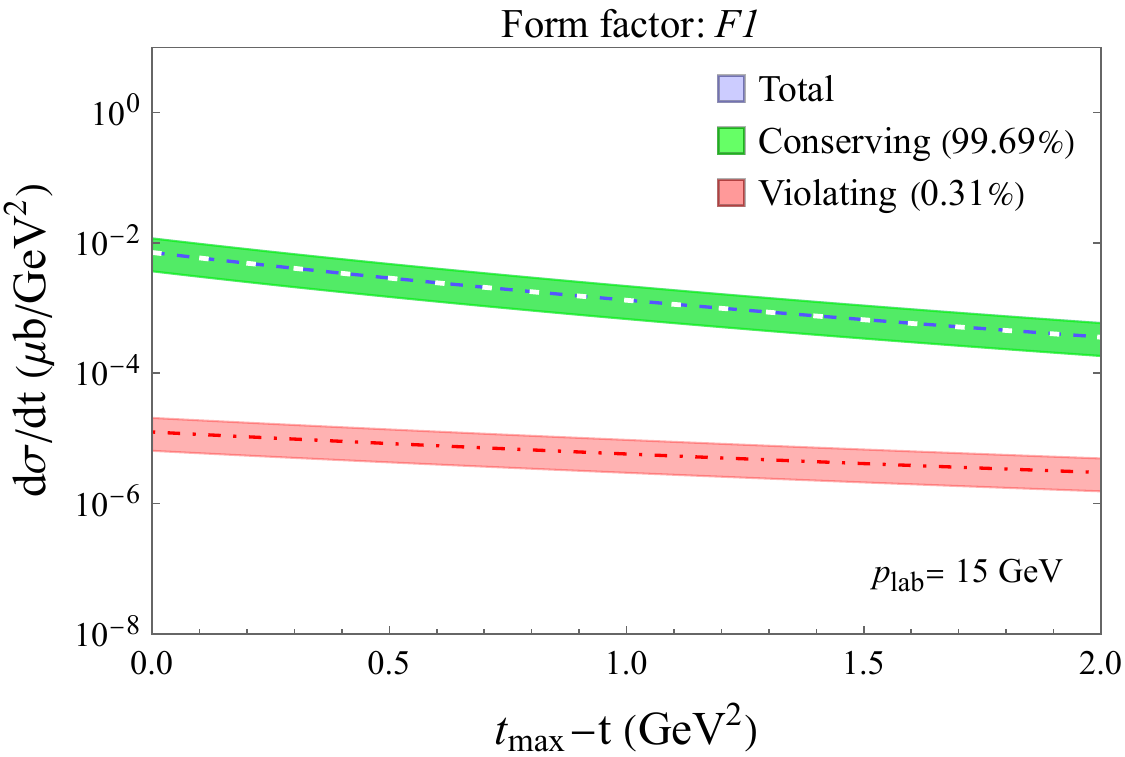}
        \caption{}
        \label{fig:ScsSc F1}
    \end{subfigure}
    \hfill
    \begin{subfigure}{0.328\textwidth}
        \centering
        \includegraphics[width=\linewidth]{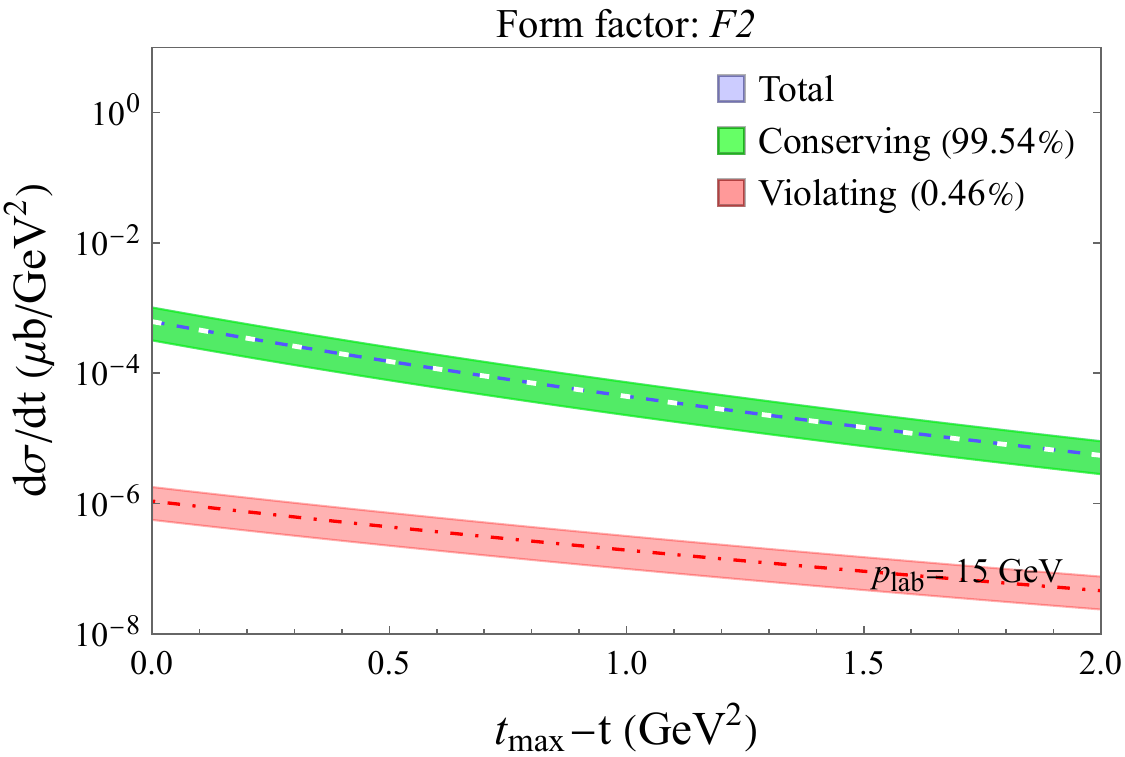}
        \caption{}
        \label{fig:ScsSc F2}
    \end{subfigure}
 \caption{Differential cross-section for $p\bar p\to\Sigma_c^*\bar \Sigma_c$ in term of conserving and violating contributions at 15 GeV.}
    \label{fig:ScsSc_combined}
\end{figure}
\begin{figure}[h]
    \centering
    \begin{subfigure}{0.328\textwidth}
        \centering
        \includegraphics[width=\linewidth]{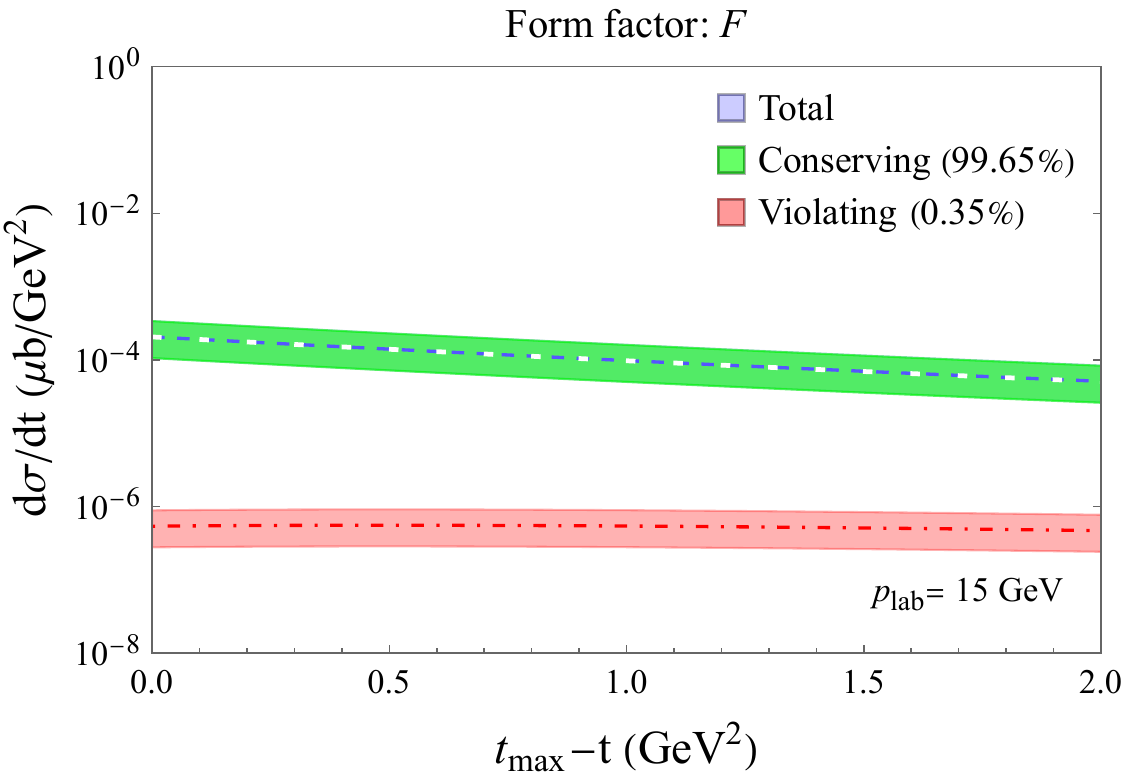}
        \caption{}
        \label{fig:ScScs F0}
    \end{subfigure}
    \hfill
    \begin{subfigure}{0.328\textwidth}
        \centering
        \includegraphics[width=\linewidth]{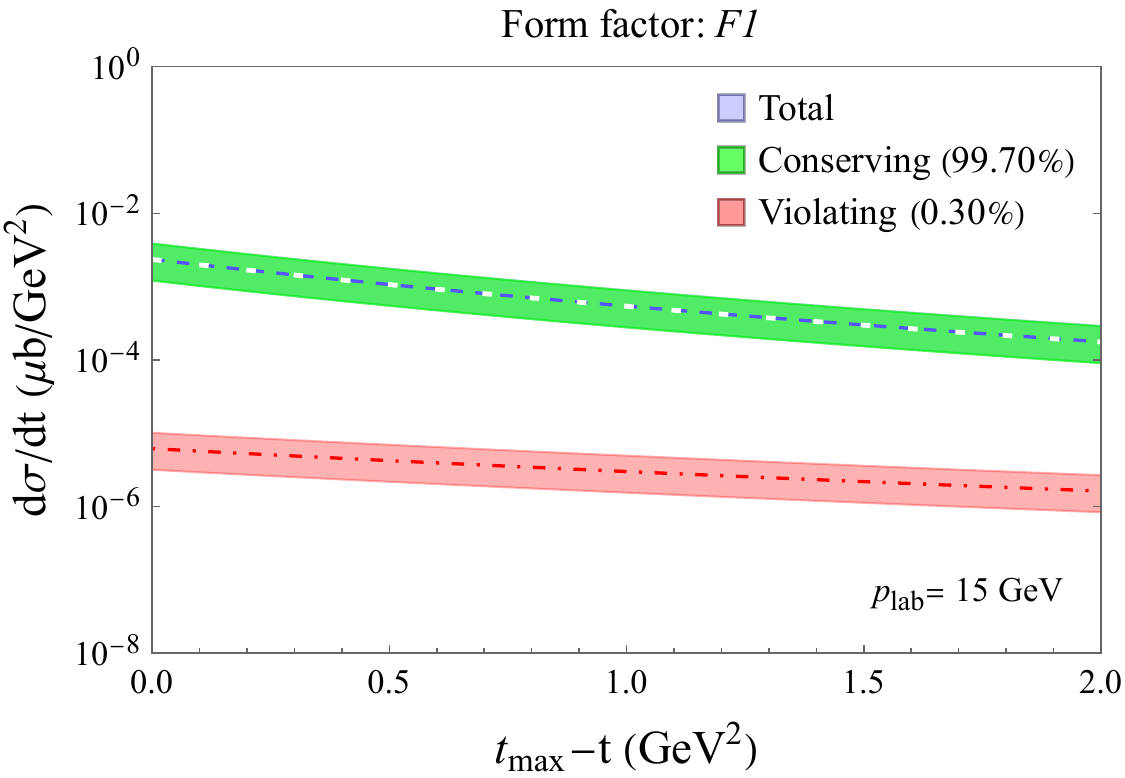}
        \caption{}
        \label{fig:ScScs F1}
    \end{subfigure}
    \hfill
    \begin{subfigure}{0.328\textwidth}
        \centering
        \includegraphics[width=\linewidth]{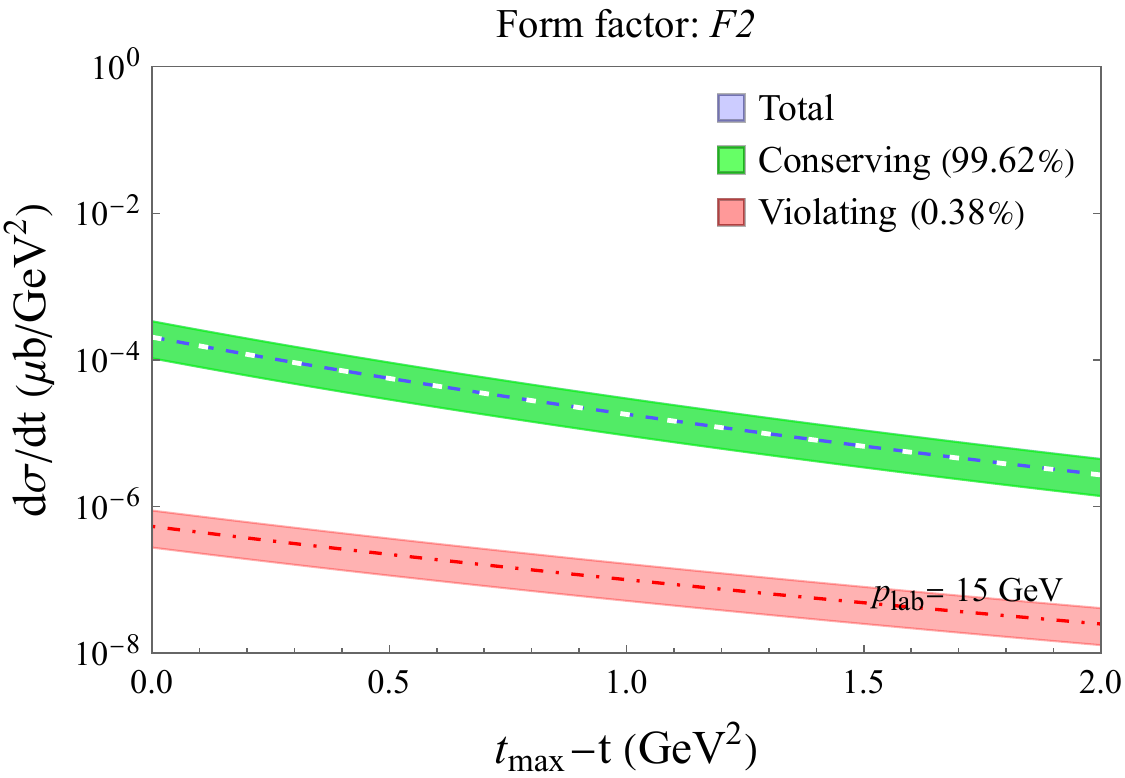}
        \caption{}
        \label{fig:ScScs F2}
    \end{subfigure}
    \caption{Differential cross-section for $p\bar p\to\Sigma_c\bar \Sigma_c^{*}$ in term of conserving and violating contributions at 15 GeV.}
    \label{fig:ScScs_combined}
\end{figure}
\begin{figure}[h]
    \centering
    \begin{subfigure}{0.328\textwidth}
        \centering
        \includegraphics[width=\linewidth]{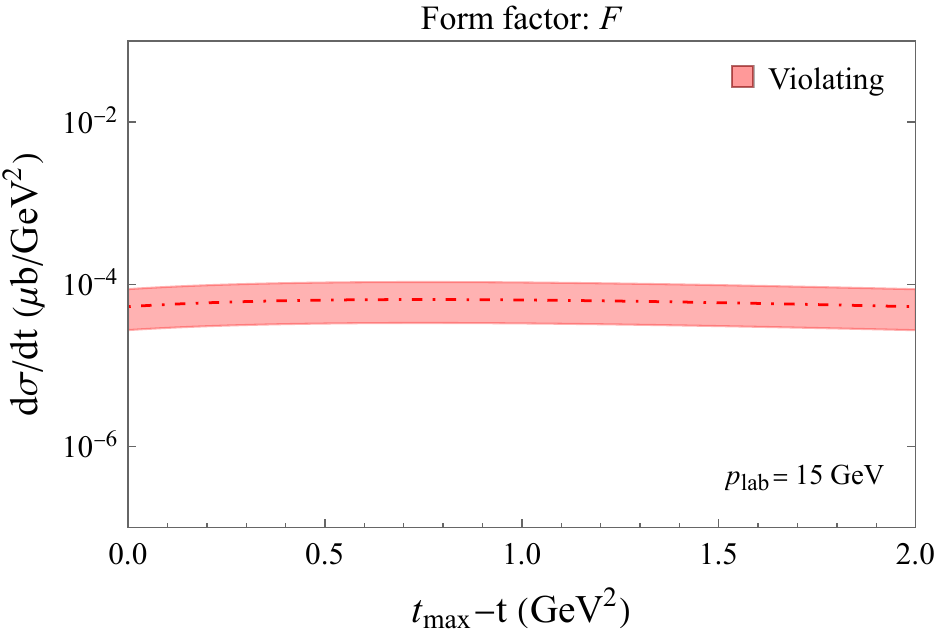}
        \caption{}
        \label{fig:LcSc F0}
    \end{subfigure}
    \hfill
    \begin{subfigure}{0.328\textwidth}
        \centering
        \includegraphics[width=\linewidth]{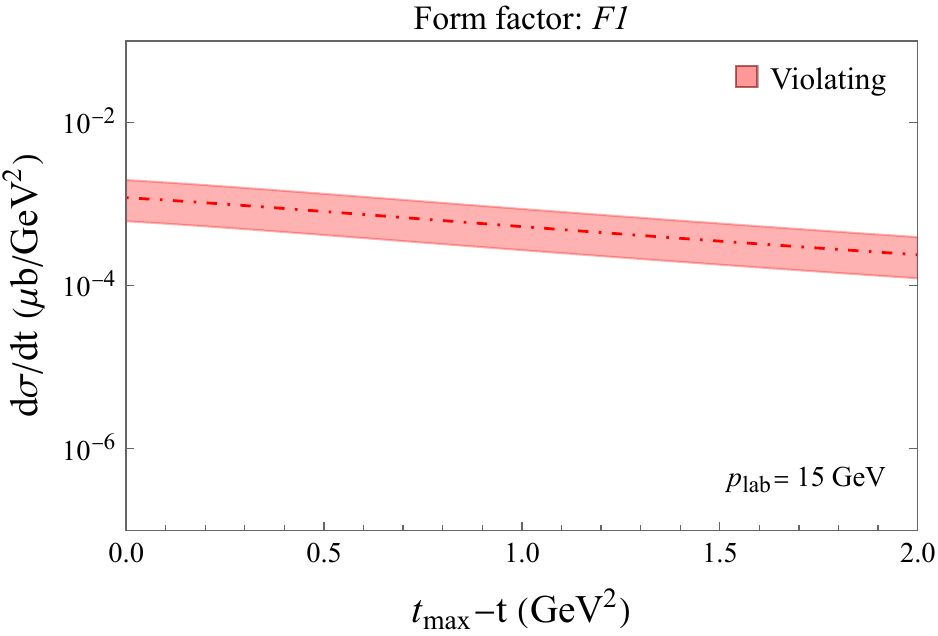}
        \caption{}
        \label{fig:LcSc F1}
    \end{subfigure}
    \hfill
    \begin{subfigure}{0.328\textwidth}
        \centering
        \includegraphics[width=\linewidth]{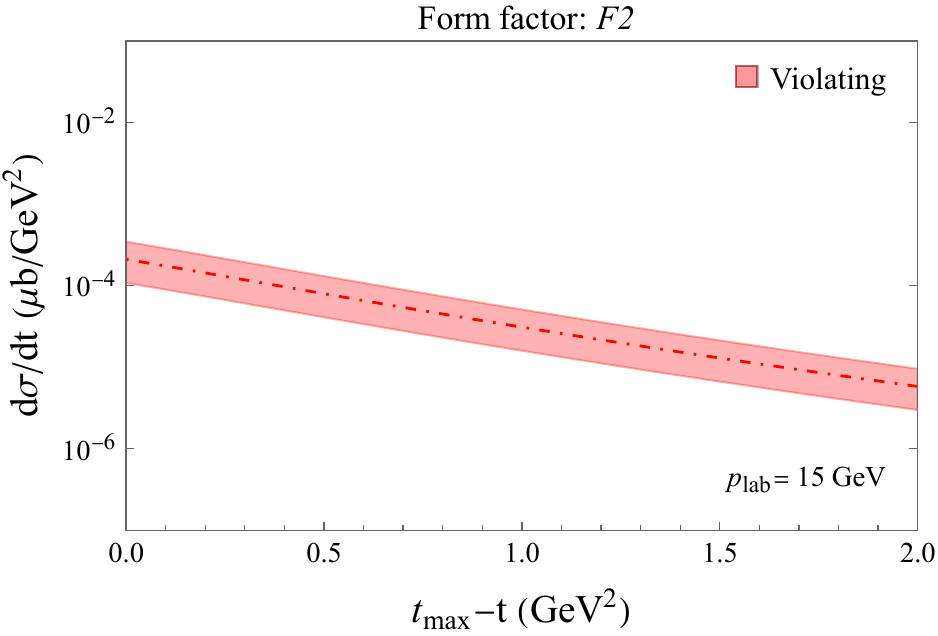}
        \caption{}
        \label{fig:LcSc F2}
    \end{subfigure}
     \caption{Differential cross-section for $p\bar p\to \Lambda_c\bar\Sigma_c$ in term of conserving and violating contributions at 15 GeV.}
    \label{fig:LcSc_combined}
\end{figure}
\begin{figure}[h]
    \centering
    \begin{subfigure}{0.328\textwidth}
        \centering
        \includegraphics[width=\linewidth]{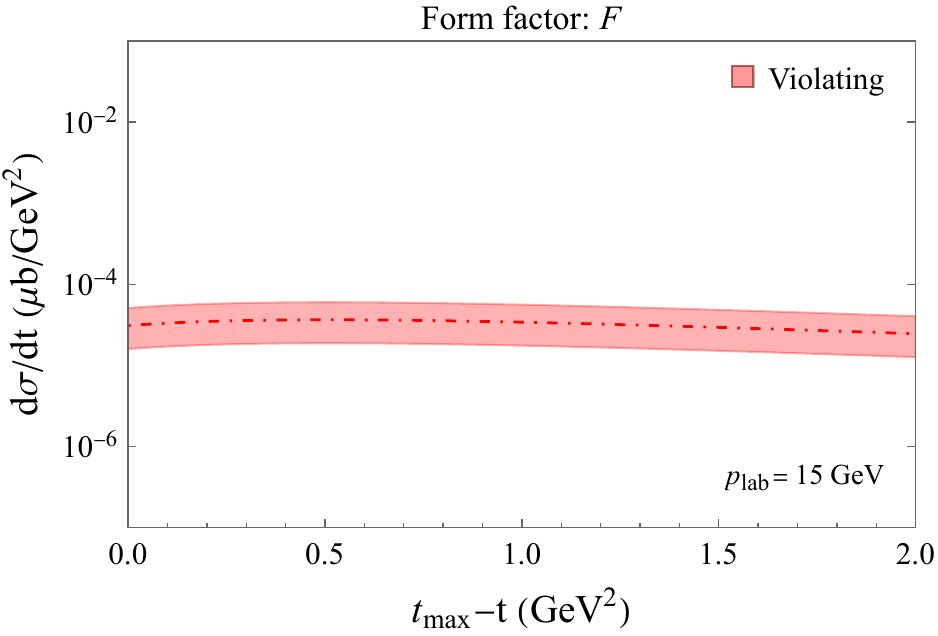}
        \caption{}
        \label{fig:ScLc F0}
    \end{subfigure}
    \hfill
    \begin{subfigure}{0.328\textwidth}
        \centering
        \includegraphics[width=\linewidth]{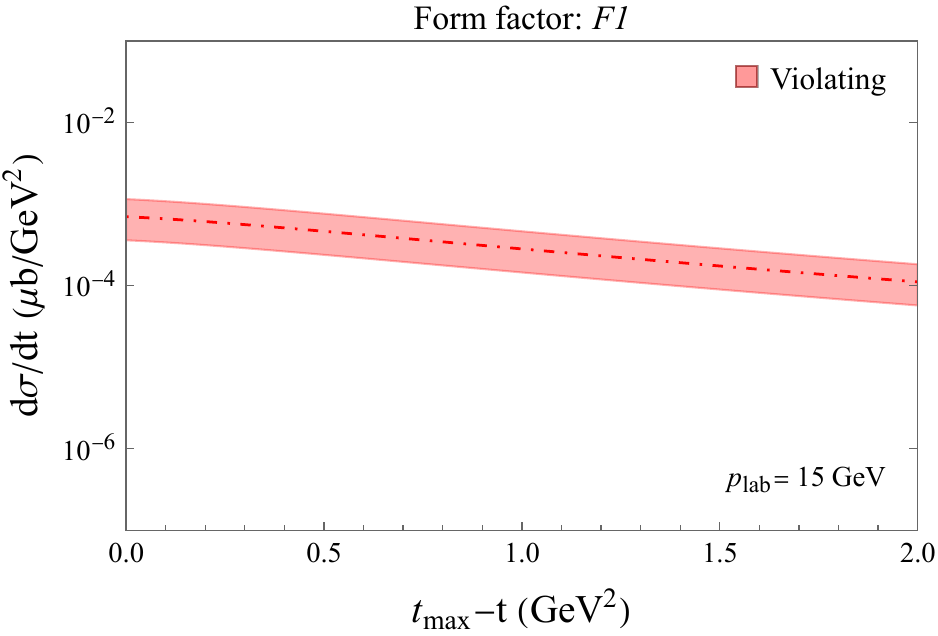}
        \caption{}
        \label{fig:ScLc F1}
    \end{subfigure}
    \hfill
    \begin{subfigure}{0.328\textwidth}
        \centering
        \includegraphics[width=\linewidth]{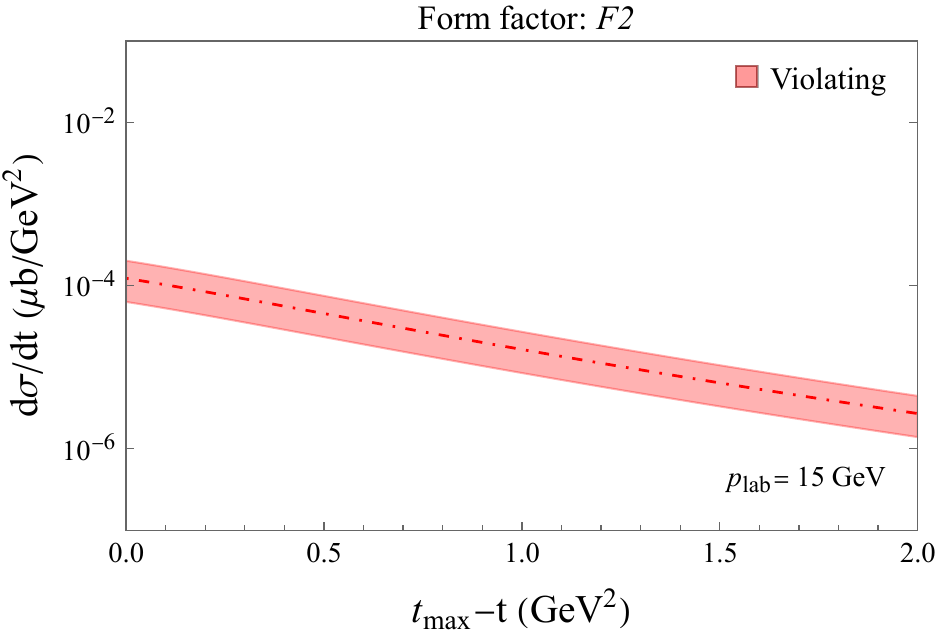}
        \caption{}
        \label{fig:ScLc F2}
    \end{subfigure}
     \caption{Differential cross-section for $p\bar p\to \Sigma_c\bar\Lambda_c$ in term of conserving and violating contributions at 15 GeV.}
    \label{fig:ScLc_combined}
\end{figure}
\begin{figure}[h]
    \centering
    \begin{subfigure}{0.328\textwidth}
        \centering
        \includegraphics[width=\linewidth]{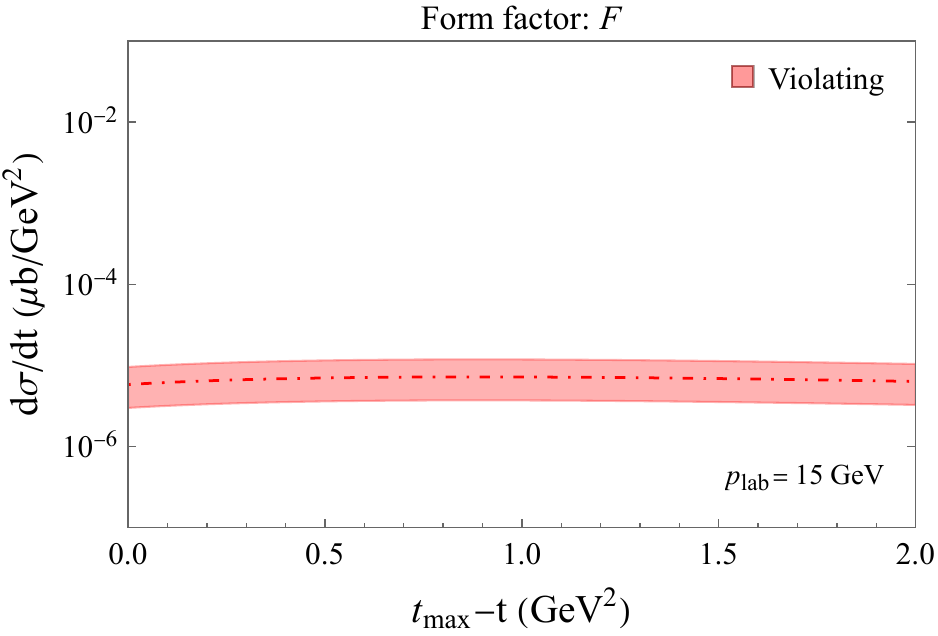}
        \caption{}
        \label{fig:LcScs F0}
    \end{subfigure}
    \hfill
    \begin{subfigure}{0.328\textwidth}
        \centering
        \includegraphics[width=\linewidth]{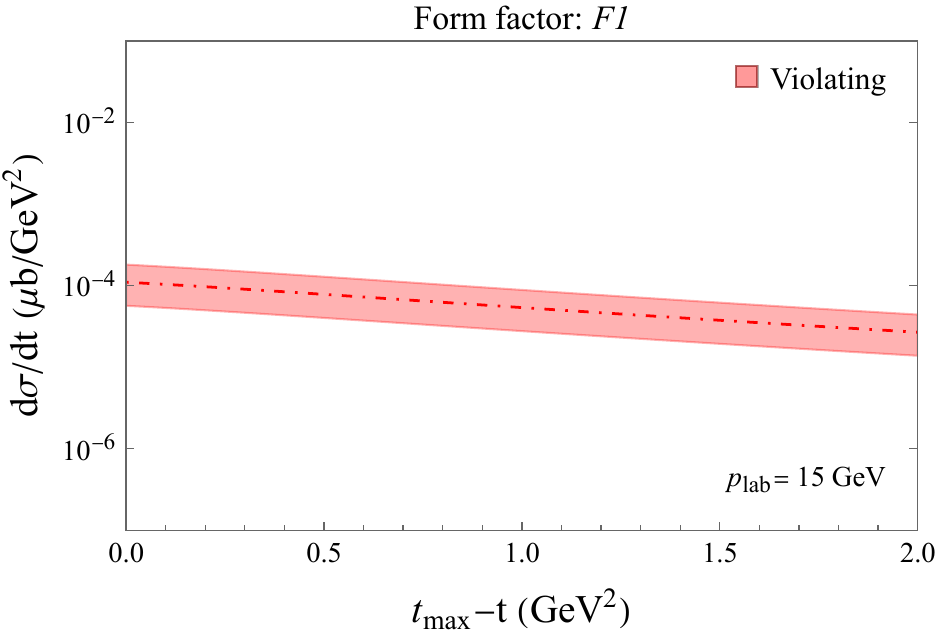}
        \caption{}
        \label{fig:LcScs F1}
    \end{subfigure}
    \hfill
    \begin{subfigure}{0.328\textwidth}
        \centering
        \includegraphics[width=\linewidth]{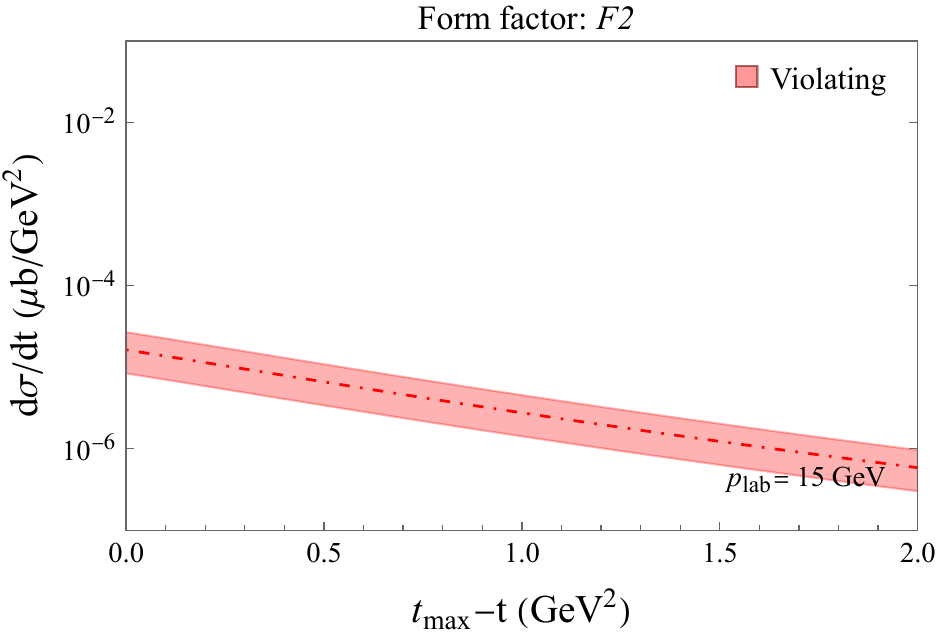}
        \caption{}
        \label{fig:LcScs F2}
    \end{subfigure}
     \caption{Differential cross-section for $p\bar p\to \Lambda_c\bar\Sigma_c^{*}$ in term of conserving and violating contributions at 15 GeV.}
    \label{fig:LcScs_combined}
\end{figure}
\begin{figure}[h]
    \centering
    \begin{subfigure}{0.328\textwidth}
        \centering
        \includegraphics[width=\linewidth]{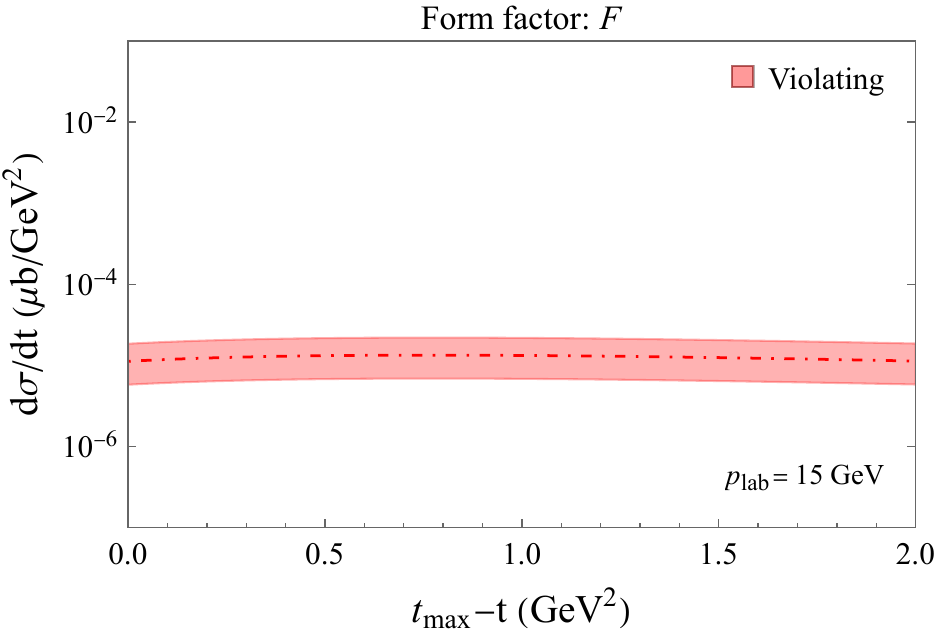}
        \caption{}
        \label{fig:ScsLc F0}
    \end{subfigure}
    \hfill
    \begin{subfigure}{0.328\textwidth}
        \centering
        \includegraphics[width=\linewidth]{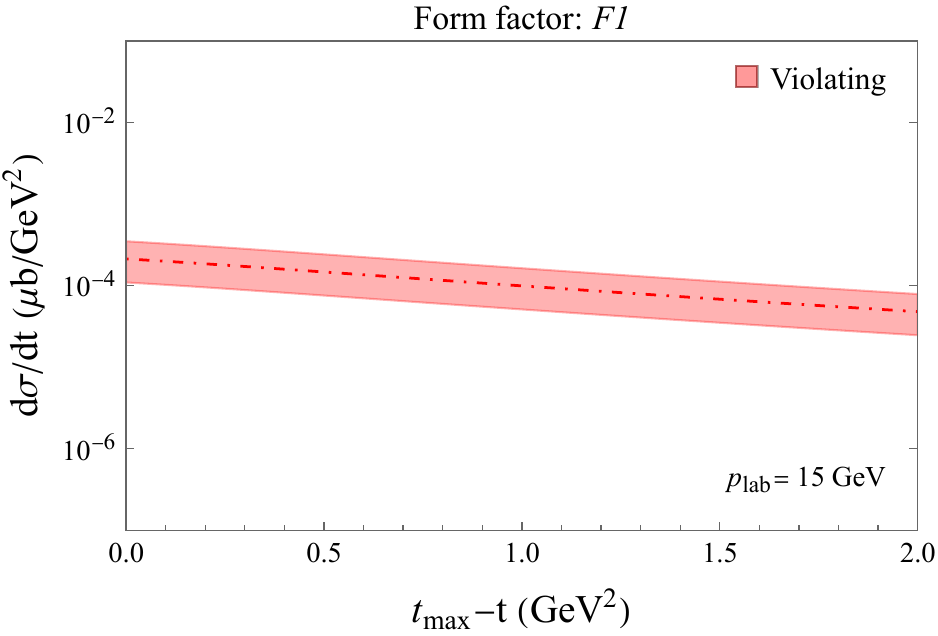}
        \caption{}
        \label{fig:ScsLc F1}
    \end{subfigure}
    \hfill
    \begin{subfigure}{0.328\textwidth}
        \centering
        \includegraphics[width=\linewidth]{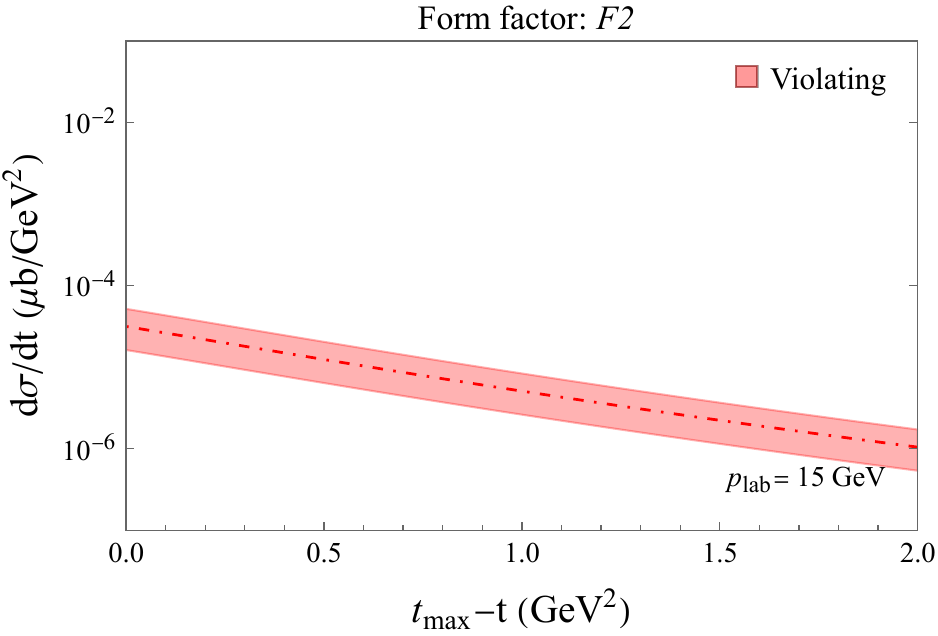}
        \caption{}
        \label{fig:ScsLc F2}
    \end{subfigure}
    \caption{Differential cross-section for $p\bar p\to \Sigma_c^*\bar\Lambda_c$ in term of conserving and violating contributions at 15 GeV.}
    \label{fig:ScsLc_combined}
\end{figure}

\FloatBarrier
\vspace{-3mm}
In~\cref{figLcLc_combined}, our predicted differential cross-sections for the reaction \( \bar{p}p \rightarrow \bar{\Lambda}_c \Lambda_c \) are presented. These contributions primarily come from the conserving terms, accounting for approximately 70--80\%, which are 80.12\%, 74.94\% and 72.45\%\,for the form factor $F$, $F_1$ and $F_2$ respectively. In contrast, the violating contributions are 19.88\%, 25.06\% and 27.55\% respectively. 
The tendencies illustrate that the exact value depends on the type of form factor. 
In figs.~\ref{fig:LcLc CF1} and \ref{fig:LcLc CF2}, the total $d\sigma/dt$ follows a similar trend in order of $10^{-1}~\mu\text{b/GeV}^2$. Meanwhile, the results in ~\cref{fig:LcLc CF0} are significantly suppressed in the order of $10^{-2}~\mu\text{b/GeV}^2$. The $d\sigma/dt$ range from \( 10^{-2} \) to \( 10^{-1}~\mu\text{b/GeV}^2 \), consistent with the findings in Ref.~\cite{Titov:2008yf}, which employs a modified Regge model inspired by quark-gluon string dynamics, with unknown parameters determined from independent studies of open strangeness production and $SU(4)_f$ symmetry. Similar results are also reported in Ref.~\cite{Khodjamirian:2011sp}, which uses Kaidalov's QGSM with Regge poles and strong couplings derived from QCD light-cone sum rules. Moreover, it indicates that the form factor \( F \) strongly suppresses the production rates. This behavior is notably consistent with other cases shown in~\cref{fig:LcSc_combined,fig:LcScs_combined,fig:ScLc_combined,fig:ScSc_combined,fig:ScScs_combined,fig:ScsLc_combined,fig:ScsSc_combined,fig:ScsScs_combined}.

In~\cref{fig:ScSc_combined}, the differential cross-section for $p\bar{p} \to \Sigma_c \bar{\Sigma}_c$ is shown. Most of the production rate is dominated by the conserving terms, contributing about $85\%-88\%$, which are $86.25\%$, $88.81\%$, and $85.13\%$ for the respective form factors, while the violating contributions are $13.75\%$, $11.19\%$, and $14.87\%$, respectively. we obtain that the conserving terms around $10^{-3} - 10^{-4}~\mu\text{b/GeV}^2$ and the violating contributions in the range of $10^{-4} - 10^{-5}~\mu\text{b/GeV}^2$. We can clearly observe that these contributions are lower than those for $p\bar{p} \to \Lambda_c \bar{\Lambda}_c$, by about two orders of magnitude ($10^{2}-10^{3}$). We found that the conserving contributions are significantly close to the studies in Refs.~\cite{Titov:2008yf,Khodjamirian:2011sp}, which estimated their order to be approximately $10^{-3}\, \mu\text{b/GeV}^2$.
\vspace{-3mm}
\FloatBarrier
\raggedbottom
Currently, the production rates of $ \Sigma_c^* \bar{\Sigma}_c^* $ have not been extensively investigated, including in experiments. Since we are performing calculations at high beam momenta $p_{\text{Lab}}= 15 \, \text{GeV} $, this leads us to further explore the spectroscopy of this production and its interactions in experiments. Our results for $ p\bar{p} \to \Sigma_c^* \bar{\Sigma}_c^* $ are shown in~\cref{fig:ScsScs_combined}. The conserving part is estimated to be about $ 100\% $, with the violation being $ 0\% $ for each form factor. The order of the conserving production rates are $ 10^{-2} - 10^{-1}~\mu\text{b/GeV}^2 $, while violation terms range from $ 10^{-7}-10^{-6}~\mu \text{b/GeV}^2 $. 
The suppression of the production rate becomes more pronounced at larger values of $t$ when using $F_2$, whereas the suppression calculated with $F$ is weaker, similar to previous results.

\vspace{-3mm}
\vspace{-3mm}

\FloatBarrier
\vspace{-3mm}
Shown in~\cref{fig:ScsSc_combined} are the results for  $p\bar{p} \to \Sigma_c^* \bar{\Sigma}_c$, which has not been widely studied either, similar to $ \Sigma_c^* \bar{\Sigma}_c^* $.
We found that the \emph{conserving} terms still play a dominant role. 
The conserving terms computed using form factors $F$, $F_1$, and $F_2$ are estimated to contribute approximately 99\%.
The differential cross-sections vary from $ 10^{-4} - 10^{-3}~\mu \text{b/GeV}^2 $. Notably, the contribution terms are slightly lower than those for $ \Sigma_c^* \bar{\Sigma}_c^* $, by about a factor of 10. When compared with $\Sigma_c \bar{\Sigma}_c^{*}$, it is moderately lower, as shown in~\cref{fig:ScScs_combined}. The production rates are in the range of $10^{-4} - 10^{-3} \, \mu \text{b/GeV}^2$ for the conserving ranges. Conversely, the violation effects fluctuate from $10^{-6}$ to nearly $10^{-5} \, \mu \text{b/GeV}^2$. Furthermore, both graphs exhibit a similar tendency. We especially observe that at $t_{\text{max}} - t = 2 \, \text{GeV}^2$, these magnitudes lie around $10^{-8} \, \mu \text{b/GeV}^2$.

%\newpage
In the following results, we present processes computed solely for violating effects, as \(\Lambda_{c}\) does not have a spin partner. In contrast, \(\Sigma_{c}\) and \(\Sigma_{c}^{*}\) are spin-coupled heavy quarks. The results are shown in figs.~\ref{fig:LcSc_combined}, \ref{fig:ScLc_combined}, \ref{fig:LcScs_combined}, and \ref{fig:ScsLc_combined}.

\vspace{-3mm}
\FloatBarrier
\raggedbottom
\vspace{-3mm}
\FloatBarrier
\raggedbottom
\vspace{-3mm}
The dynamics of \(\Lambda_c \bar{\Sigma}_c\) are less studied mainly because \(\Sigma_c \bar{\Lambda}_c\) is easier to study experimentally. The \(\Sigma_c \bar{\Lambda}_c\) channel exhibits stronger interactions and fits better with simpler theoretical models. However, to fully understand the \(p\bar{p}\) annihilation process, it is important to study both final states. 
Our prediction for \(p\bar{p} \to \Lambda_c \bar{\Sigma}_c\) with violation effects, as shown in~\cref{fig:LcSc_combined}, indicates that the production rate is approximately in the range of \(10^{-4} - 10^{-3} \, \mu  \text{b/GeV}^2\), with small differences among the form factors. \(F_1\) consistently shows the highest peak, while \(F\) initially appears to be the lowest. 
However, \(F_2\) becomes more suppressed as \(t\) increases, in agreement with previous calculations.
Compared to \(\Sigma_c \bar{\Lambda}_c\) in~\cref{fig:ScLc_combined}, the range remains similar, though slightly reduced.
\vspace{-3mm}
\FloatBarrier
\raggedbottom
\vspace{-3mm}
\FloatBarrier
\vspace{-3mm}
The production rates for the charmed baryon triplet \(\Sigma_c^*\) (\(J^P = 3/2^+\)) and the charmed baryon singlet \(\Lambda_c\) (\(J^P = 1/2^+\)) in the processes \( p\bar{p} \to \Sigma_c^*\bar{\Lambda}_c, \Lambda_c\bar{\Sigma}_c^* \) are computed, as depicted in figs.~\ref{fig:LcScs_combined} and~\ref{fig:ScsLc_combined}. The production rate falls within the interval \( 10^{-4} - 10^{-5} \, \mu \text{b/GeV}^2 \), with only minor variations. 
In addition, it shows an increase by a factor of \(10\) compared to \(\Sigma_{c}^{*}\bar{\Sigma}_c\) and \(\Sigma_{c}\bar{\Sigma}_{c}^{*}\), and are roughly \( 100 \) greater than \(\Sigma_{c}^{*}\bar{\Sigma}_{c}^{*}\).
%\vspace{-3mm}
\subsection{Numerical results of the charmed baryon productions}
%\vspace{-3mm}
In this subsection, we calculate the total cross-sections, \(\sigma_{\text{total}}\), and compare with previous studies, focusing on different \(p_{\text{Lab}} (\text{GeV}/c)\) thresholds, as illustrated in Table~\ref{tab:cross_sections}.

\begin{table}[h]
\centering
\resizebox{\textwidth}{!}{ % Adjust to fit the page width
\renewcommand{\arraystretch}{1.5} % Adjust row height for better readability
\setlength{\tabcolsep}{10pt} % Adjust column spacing
\begin{tabular}{| c | c | c | c | c | c | c |} 
\hline
\multirow{2}{*}{\Large${p\bar{p} \to Y_c\Bar{Y}'_c}$} 
& \multicolumn{6}{c|}{\large{Total Cross-Section (\Large\(\sigma_{\text{total}}\)) (\(\mu \text{b}\))}} \\ 
\cline{2-7}
&\,\,\,\,\,\,\,\,\,\,\,\,\,\large\text{QGSM}\,\cite{Kaidalov:1994mda}\,\,\,\,\,\,\,\,\,\,\,\,\, & \large\text{Handbag approach}\,\cite{Goritschnig:2009sq} & \large\text{Quark-gluon dynamics \cite{Krein:2010zza,Haidenbauer:2010nx}} & \large\text{Kaidalov's QGSM \cite{Khodjamirian:2011sp}} & \large\text{Quark-gluon dynamics} \cite{Haidenbauer:2016pva} & \large\text{This study} \\ 
\hline \hline
\multirow{3}{*}{\Large$\Lambda_c\bar{\Lambda}_c$} 
& \multirow{3}{*}{\large$1.03\times10^{-2}$} 
& \multirow{3}{*}{\large$(5.88-9.92)\times10^{-4}$} 
& \multirow{3}{*}{\large$(2.90-19.5)\times10^{-1}$} 
& \multirow{3}{*}{\large$(1.98-28.2)\times10^{-2}$} 
& \multirow{3}{*}{\large$1.94-3.07$} 
& \large$(0.940-2.79)\times10^{-2}$ ($F_0$) \\ 
&&&&&& \large$(2.65-7.89)\times10^{-2}$ ($F_1$) \\
\Large$(p_{\text{Lab}} = 10.5 \, \text{GeV}/c)$ 
&&\Large$(p_{\text{Lab}} = 10.7 \, \text{GeV}/c)$&&&& \large$(0.543-1.61)\times10^{-3}$\,($F_2$) \\
\hline
\multirow{3}{*}{\Large$\Sigma_c\bar{\Sigma}_c$} 
& \multirow{3}{*}{\large-} 
& \multirow{3}{*}{\large-} 
& \multirow{3}{*}{\large-} 
& \multirow{3}{*}{\large$(1.07-60.2)\times10^{-3}$} 
& \multirow{3}{*}{\large$(4.92-9.68)\times10^{-4}$} 
& \large$(2.02-6.28)\times10^{-5}$\,\,($F_0$) \\
&&&&&& \large$(5.04-15.8)\times10^{-5}$\,\,($F_1$) \\
\Large$(p_{\text{Lab}} = 12 \, \text{GeV}/c)$ 
&&&&&& \large$(4.64-14.5)\times10^{-7}$\,\,($F_2$) \\
\hline
%& & & & & & {Only VHQSS} \\
\multirow{2}{*}{\Large$\Sigma_c\bar{\Lambda}_c$} 
& \multirow{2}{*}{\large-} 
& \multirow{2}{*}{\large-} 
& \multirow{2}{*}{\large-} 
& \multirow{2}{*}{\large$(5.05-136)\times10^{-3}$} 
& \multirow{2}{*}{\large$(5.09-9.83)\times10^{-3}$} 
& \large$(6.95-22.2)\times10^{-6}$\,\,($F_0$) \\
&&&&&& \large$(1.92-6.14)\times10^{-5}$\,\,($F_1$) \\
\Large$(p_{\text{Lab}} = 11 \, \text{GeV}/c)$ 
&&&&&& \large$(1.93-6.18)\times10^{-7}$\,\,($F_2$) \\
\hline
\multirow{3}{*}{\Large$\Sigma_c^{*}\bar{\Sigma}_c^{*}$} 
& \multirow{3}{*}{\large-} 
& \multirow{3}{*}{\large-} 
& \multirow{3}{*}{\large-} 
& \multirow{3}{*}{\large-} 
& \multirow{3}{*}{\large-} 
& \large$(4.02-12.9)\times10^{-2}$\,\,($F_0$) \\
&&&&&& \large$(1.32-4.21)\times10^{-1}$\,\,($F_1$) \\
\Large$(p_{\text{Lab}} = 15 \, \text{GeV}/c)$ 
&&&&&& \large$(3.81-12.2)\times10^{-3}$\,\,($F_2$) \\
\hline
\hline
\end{tabular}
}
\caption{Total cross-sections (\(\sigma_{\text{total}}\)) for different channels in the \(p\bar{p} \to Y_c\Bar{Y}'_c\) process.}
\label{tab:cross_sections}
\end{table}

%\FloatBarrier  % Forces placement of all floats before continuing
%\newpage%

Our predicted total cross-sections for the reaction \(\Lambda_c \bar{\Lambda}_c\), for each form factor at \(p_{\text{Lab}} = 10.5\, \text{GeV/c}\), are in the range of \(10^{-3} - 10^{-2}\,\mu \text{b}\). These values are consistent with the range reported in Ref.~\cite{Kaidalov:1994mda}, predicated on the non-perturbative QGSM, and are in approximate agreement with Ref.~\cite{Khodjamirian:2011sp}, which employs Kaidalov's QGSM with Regge poles and uses strong couplings derived from QCD light-cone sum rules. However, they are about 10 times smaller than the results in Refs.~\cite{Krein:2010zza,Haidenbauer:2010nx}, which are predicated on a meson-exchange framework and use $SU(4)_f$ symmetry to fix coupling constants, and approximately 100 times smaller than those from investigations of kinematic thresholds predicated on quark-gluon dynamics \cite{Haidenbauer:2016pva}. For \(F_2\), the cross-sections deviate significantly, being more than 10 times smaller than those for the other form factors. However, they are higher by a factor of 10 at \(p_{\text{Lab}} = 10.7 \, \text{GeV}/c\), as roughly evaluated in Ref.~\cite{Goritschnig:2009sq}.

For $\Sigma_{c}\bar{\Sigma}_{c}$ production at \(p_{\text{Lab}} = 12 \, \text{GeV}/c\). We find that the cross-sections fall in the range of $10^{-7} - 10^{-5}\,\mu \text{b}$, which decreases dramatically compared to $\Lambda_{c}\bar{\Lambda}_{c}$, by about $10^{3} - 10^{4}$ times. Similarly, in Ref.~\cite{Haidenbauer:2016pva}, the $\Lambda_{c}\bar{\Lambda}_{c}$ production is higher by a factor of approximately $10^{4}$. Additionally, our results are less than those in Ref.~\cite{Khodjamirian:2011sp}, on the order of $10^{-2}\,\mu \text{b}$ for the form factors $F$ and $F_1$, and approximately $10^{-4}\,\mu \text{b}$ for $F_2$, which is significantly more suppressed.

For  \(\Sigma_{c}\Bar{\Lambda}_{c} \) production at a laboratory momentum of \( 11\,\text{GeV}/c \), only the violating contribution is calculated, with values ranging between \( 10^{-5} \) and \( 10^{-7} \, \mu \text{b} \). These results are smaller compared to those found in Refs.~\cite{Khodjamirian:2011sp, Haidenbauer:2016pva}, which are approximately \( 10^{-3} \, \mu \text{b} \).

Finally, We have predicted  the total cross-sections for \( p\Bar{p} \rightarrow \Sigma_{c}^{*}\Bar{\Sigma}_{c}^{*} \) at \(p_{\text{Lab}} = 15 \, \text{GeV}/c\), which has not been extensively studied, with values of $(4.02-12.9)\times10^{-2}~\mu \text{b} $\,($F_0$), 
$(1.32-4.21)\times10^{-1}~\mu \text{b} $\,($F_1$) 
and $(3.81-12.2)\times10^{-3}~\mu \text{b} $ \,($F_2$).
%\FloatBarrier
%\clearpage
\section{Conclusions}\label{sec-5}
We constructed the $SU(2)_f$ effective Lagrangians with 12 LECs. The HQSS-invariant Lagrangian reveals that pseudoscalar \(D\) meson couple to nucleons and \(\Lambda_c\) baryons, while vector \(D\) meson couple to nucleons and \(\Sigma_c\) (\(\Sigma_c^*\)) baryons. Including heavy-quark spin symmetry violations recovers all interactions with 9 residual LECs. Moreover, we investigated \(d\sigma/dt\) for charmed production processes in $p\bar{p}$ collisions and evaluated the contributions from HQSS and its violation. As a result, the conserving contributions dominate at 70-80\% for \( \Lambda_{c}\bar{\Lambda}_{c} \), 85-86\% for \( \Sigma_{c}\bar{\Sigma}_{c} \), and about 99\% for \( \Sigma_{c}^{*}\bar{\Sigma}_{c} \) and \( \Sigma_{c}\bar{\Sigma}_{c}^{*} \), with \( \Sigma_{c}^{*}\bar{\Sigma}_{c}^{*} \) being the most dominant at around 100\%. These results indicate that \emph{conserving} HQSS works well when the particles have heavy-quark spin partners, while more  HQSS breaking effects appear in \( \Lambda_{c}\bar{\Lambda}_{c} \).

The total \( d\sigma/dt \) for \( p\bar{p} \rightarrow \Lambda_{c}\bar{\Lambda}_{c} \) and \(\Sigma_{c}\bar{\Sigma}_{c} \), vary between \( 10^{-2}-10^{-1}~\mu\text{b/GeV}^2 \) and \( 10^{-4}-10^{-3}~ \mu\text{b/GeV}^2 \) respectively,  which is consistent with the results reported in Refs.~\cite{Titov:2008yf,Khodjamirian:2011sp}. Although the total \( d\sigma/dt \) for \( \Sigma_{c}^{*}\bar{\Sigma}_{c}^{*} \) and \( \Sigma_{c}^{*}\bar{\Sigma}_{c}\,(\Sigma_{c}\bar{\Sigma}_{c}^{*}) \) are significantly greater than \( \Sigma_{c}\bar{\Sigma}_{c} \) by factors of about 100 and 10, respectively, the violated terms dramatically decrease, by 100 times and 10 times. Additionally, they indicate that as \( t \) increases, the form factor \( F_2 \) induces a stronger suppression. 

We also presented the predicted $\sigma_{\text{total}}$ for $\Lambda_{c}\Bar{\Lambda}_c$ ($p_{\text{Lab}} = 10.5 \, \text{GeV}/c)$ estimated to be in the range of $10^{-2} - 10^{-3}~\mu \text{b}$, in close agreement with previous studies in Refs.~\cite{Kaidalov:1994mda,Khodjamirian:2011sp}, with only slight differences compared to the results reported in Refs.~\cite{Goritschnig:2009sq,Krein:2010zza,Haidenbauer:2010nx}. For $\Sigma_{c}\Bar{\Sigma}_c$ ($p_{\text{Lab}}=12 \, \text{GeV}/c$), the constrained form factors $F_{0}$ and $F_{1}$ are lower by about 10 to 100 times compared to those in Refs.~\cite{Khodjamirian:2011sp,Haidenbauer:2016pva}. More importantly, we computed $\sigma_{total}$ for $\Sigma^{*}_{c}\Bar{\Sigma}^{*}_{c}$ ($p_{\text{Lab}}=15 \, \text{GeV}/c$), with results aligning in the range of $(4.02-12.9) \times 10^{-2}~\mu \text{b}$ ($F_0$), $(1.32-4.21) \times 10^{-1}~\mu \text{b} $ ($F_1$), and $(3.81-12.2) \times 10^{-3}~\mu \text{b}$ ($F_2$).

For the forthcoming $\bar{\text{P}}$ANDA experiments at FAIR, the High-Energy Storage Ring (HESR) will store antiprotons in a momentum range from 1.5 to 15 GeV/c \cite{Frankfurt:2019uvk,PANDA:2009yku}. Hopefully, our results will provide in exploring the nature of charmed baryons and also serve as the first step towards more involved reaction mechanisms, leading to an increase in experimental requirements.

\section*{Acknowledgement}
N.M. is financially supported by Development and Promotion of Science and Technology Talents Project (DPST), Thai government scholarship.
N.P. is financially supported by the National Astronomical Research Institute of Thailand (NARIT). A.J.A. was supported by RIKEN special postdoctoral researcher program. D.S. is supported by the Fundamental Fund of Khon Kaen University and has received funding support from the National Science, Research and Innovation Fund and supported by Thailand NSRF via PMU-B [grant number B39G670016]. 

\bibliographystyle{ieeetr} %
\bibliography{biblio}
\end{document}